\newif\ifShowKeys
\ShowKeystrue

\documentclass[10pt,a4paper]{article}
\usepackage[no-natbib-sort]{jheppub}
\usepackage[utf8]{inputenc}
\usepackage[english]{babel}
\usepackage{comment}
\usepackage{MnSymbol}
\usepackage{mathtools}
\usepackage{environ}
\usepackage{amsthm}
\usepackage[normalem]{ulem}
\usepackage{cancel}
\usepackage[dvipsnames]{xcolor}

\newcommand{\mR}{\ensuremath{\mathbb R}}
\newcommand{\mN}{\ensuremath{\mathbb N}}

\newcommand{\mE}[1]{\ensuremath{\mathbb E_{#1}}}
\newcommand{\Bav}[2]{\ensuremath{\omega_{#2}(#1)}}
\newcommand{\rBav}[3]{\ensuremath{\Omega_{#2}^{(#3)}(#1)}}
\newcommand{\QBav}[2]{\ensuremath{\langle #1\rangle_{#2}}}
\newcommand{\rQBav}[3]{\ensuremath{\langle #1\rangle_{#2}^{(#3)}}}
\newcommand{\vphi}{\ensuremath{\varphi}}
\newcommand{\mP}{\ensuremath{\mathbb P}}
\newcommand{\esssup}{\ensuremath{\mathrm{ess \, sup}}}

\newcommand{\stkout}[1]{\ifmmode\text{\sout{\ensuremath{#1}}}\else\sout{#1}\fi}

\NewEnviron{es}{
	\begin{equation}\begin{split}
			\ensuremath{\BODY}
	\end{split}\end{equation}
}
\NewEnviron{ess}{
	\begin{equation*}\begin{split}
			\ensuremath{\BODY}\nonumber
	\end{split}\end{equation*}
}
\newcommand{\mc}{\mathcal}
\newcommand{\bb}[1]{\boldsymbol{#1}}
\newcommand{\red}[1]{#1}

\newtheorem{Thm}{Theorem}
\newtheorem{Cor}{Corollary}
\newtheorem{Def}{Definition}
\newtheorem{Rmk}{Remark}
\newtheorem{Prp}{Proposition}
\newtheorem{Lem}{Lemma}

\title{\red{The infinite volume limit of the Hopfield neural networks in the replica-symmetric regime}}
\author[a,1,3,5]{Elena Agliari,}
\author[b,1,3,5]{Adriano Barra,}
\author[c]{Pierluigi Bianco,}
\author[a,1]{Alberto Fachechi,}
\author[c,2,4]{Diego Pallara}

\affiliation[a]{Dipartimento di Matematica ``Guido Castelnuovo'', Sapienza Universit\`a di Roma, Italy}
\affiliation[b]{Dipartimento di Scienze di Base ed Applicate per l'Ingegneria, Sapienza Universit\`a di Roma, Italy}
\affiliation[c]{Dipartimento di Matematica e Fisica ``Ennio De Giorgi'', Universit\`a del Salento, Italy}
\affiliation[1]{Gruppo Nazionale per la Fisica Matematica (GNFM-INdAM), 
Italy}
\affiliation[2]{Istituto Nazionale di Fisica Nucleare, Sezione di Lecce, Italy}
\affiliation[3]{Istituto Nazionale di Fisica Nucleare, Sezione di Roma, Italy}
\affiliation[4]{Gruppo Nazionale per l'Analisi Matematica, Probabilità e le loro Applicazioni (GNAMPA-INdAM), Italy}
\affiliation[5]{CNR Nanotec, Salento Unit, Lecce, Italy}
\emailAdd{alberto.fachechi@uniroma1.it}
\abstract{Despite extensive research over the past five decades, a rigorous determination of the infinite-volume limit of the free energy of the Hopfield model \red{with an extensive number of patterns} has remained an outstanding open problem as standard methods from spin-glass theory have so far proven inadequate for this purpose.  While a complete proof \red{for finite, non-vanishing, loads} is still out of reach, here we develop a novel interpolation strategy that allows us to express the free energy of the Hopfield model in terms of a linear combination of the free energies of hard- and soft-spin glass models along with a remainder contribution. The payoff of this interpolation scheme is twofold: 
\newline
(i)  As the remainder involves positive-definite fluctuations of the order parameters (and it vanishes under a concentration assumption on their means where we consistently recover the replica-symmetric picture) such an expression finally constitutes a rigorous {\em bound} on the exact free energy, rather than just an {\em approximation,} \red{as it was the case for earlier interpolation procedures à la Guerra}.
\newline
(ii)  As for the spin-glass models the existence of the infinite-volume free energy has been previously determined by the Guerra--Toninelli scheme, this novel interpolation allows us to extend its validity to the Hopfield model too. Yet, as the equivalence between the Hopfield model and mixtures of  spin-glasses is currently well established under replica symmetry and up to the first step of replica symmetry breaking, at present our approach can be rigorously implemented only within these regimes. 
\newline
\red{In this work, by inspecting how the order parameters  have to self-average around their means, we restrict our attention to the replica-symmetric regime}.}

\keywords{Thermodynamic limit, neural networks, spin glasses, mean-field models}
\begin{document}
\maketitle

\section{Introduction}
\begin{figure}
    \centering
    \includegraphics[width=1.0\linewidth]{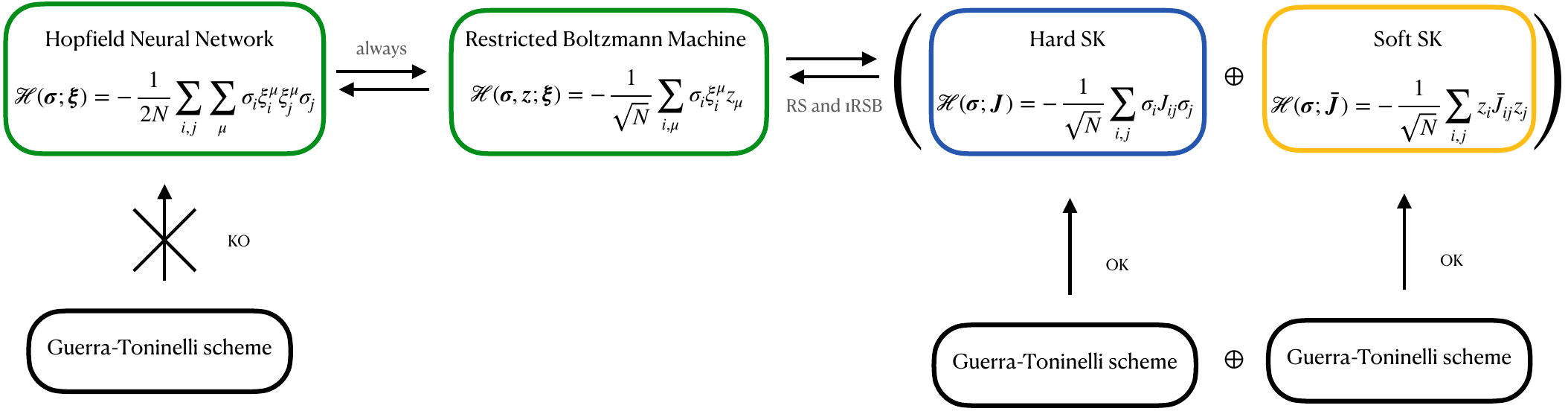}
    \caption{Structure of the proof: as the Guerra-Toninelli scheme does not apply directly to the Hopfield model, but it does on both the hard and the soft Sherrington-Kirkpatrick models, at first we map the Hopfield model into a restricted Boltzmann machine, then we use the decomposition of the quenched free energy of the latter in terms of a mixture of those pertaining to these spin glasses. However, while the first passage (from the Hopfield model to the restricted Boltzmann machine) holds generally, the second passage (from the restricted Boltzmann machine to a mixture of spin glasses), at present, is known to hold solely at the replica-symmetric level  (see Theorem $4$ of \cite{barra2012glassy})  or under one step of replica symmetry breaking (see Theorem $9$ of \cite{LindaRSB}) and, in this first work, we confine our inspection to the replica-symmetric case.}
    \label{figure-one}
\end{figure}

Since the seminal works of Amit, Gutfreund, and Sompolinsky (AGS) \cite{AGS1,AGS2}, the Hopfield model \cite{Hopfield} has become one of the paradigmatic models of associative memory and pattern recognition in neural networks (see, e.g., \cite{Lenka,AlluNeed,Lucibello,Carleo,AlboDream}). A major reason for its enduring relevance lies in the possibility of describing its collective computational behavior through the powerful analytical framework originally developed for spin glasses \cite{Amit,MPV}. Indeed, the statistical-mechanical approach pioneered in the 1980s provided a remarkably successful explanation of the model's emergent memory and retrieval capabilities.
Yet, despite this conceptual success, the mathematical understanding of neural networks remains substantially less complete than that of their spin-glass counterparts and a number of fundamental questions are still open. 
Specifically, for paradigmatic examples of spin glasses, such as the Sherrington-Kirkpatrick model \cite{sherrington1972solvable}, a full mathematical control has been reached and the Parisi theory \cite{Giorgio1,Giorgio2,Giorgio3} has been confirmed rigorously. On the other hand, despite the extensive results collected so far (see e.g. \cite{Tala,Anton1,Anton2,Anton3,Anton4,TalaNN1,TalaNN2}),  for the Hopfield model we still lack a fully comprehensive Parisi theory \red{at non-vanishing loads} \cite{MultipleEquilibria} and, actually, even a proof of the existence of the thermodynamic limit of its free energy is missing.
\newline
While such a proof is still out of reach, in these notes we aim to take a small step forward, following an approach similar in spirit to that pursued by Pastur, Shcherbina and Tirozzi \cite{PST1,PST2}, namely, we assume that the order parameters of the theory do not fluctuate in the asymptotic limit and we use this hypothesis to prove the existence of the infinite volume limit for the quenched free energy. In particular, we precisely quantify which are the moments of the order parameter distributions that must vanish asymptotically in order to obtain the AGS replica-symmetric expression \cite{AGS2}.
\newline
The plan is to leverage recent advances in our understanding of the Hopfield model to ``{circumvent}'' the obstacles that prevent  the application of the standard Guerra-Toninelli argument \cite{guerra2002thermodynamic,guerra2002central}. 
\newline
At first, we linearize the Hopfield Hamiltonian by exploiting its equivalence \cite{BarraEquivalenceRBMeAHN,ABT-RBM,Sollich2018,MonassonPRL,Auro1,Auro2} with a restricted Boltzmann machine that, in a statistical mechanical jargon, is a hybrid bipartite spin-glass \cite{Equilibrium} largely used in Machine Learning \cite{RBMoriginal}.
\newline
Next, we rely on a class of representation theorems\footnote{At present, these representation theorems are available at replica-symmetric level (see Theorem $4$ of \cite{barra2012glassy}) as well as at a broken-replica-symmetry level (see Theorem $9$ of \cite{LindaRSB} and \cite{pelatino}) but solely for the first RSB step: in the near future, further (computationally extensive but fairly standard) efforts would reasonably generalize these results to account for more RSB steps.} that allows us to recast the free energy of a restricted Boltzmann machine in terms of a linear combination of the free energies of two spin-glass models, the former being the (standard, i.e., {\em hard}) Sherrington-Kirkpatrick model \cite{sherrington1972solvable}, the latter being its {\em soft} (or spherical or Gaussian) counterpart \cite{Isaac,dembo1,dembo2}: crucially for both these models the Guerra-Toninelli argument does apply and the asymptotic limits of their free energies \red{have} been proven to exist \cite{barra2014solvable}.
\newline
Then, we introduce an interpolating free energy à la Guerra \cite{guerra2001sum,Guerra1,Fachechi1}, whose extrema return the quenched free energy of the original Hopfield model and a mixture of the quenched free energies of these spin glasses \red{with extensive volume}. By this interpolation, 
assuming self-averaging properties for the order parameters in the thermodynamic limit, the Hopfield model's quenched free energy can be estimated exactly (confined to the replica-symmetric level of description as a natural consequence of the concentration of measure assumption): the work-flow of our approach is summarized in Fig. \ref{figure-one}.  
\newline
As a side remark,  we note that by this new interpolation technique finally the replica-symmetric solution now acts a rigorous bound for the free energy and no longer just as an approximation as in previous interpolation schemes \cite{Fachechi1,barattolo,barra2012glassy,Guerra2}.  
\newline
The paper is structured as follows: in Sec. \ref{SezioneDue}, we provide the main definitions of the model(s), their related statistical mechanical formulation, as well as the main theorems on the asymptotic limits we will employ.  In Sec. \ref{SezioneTre} we show the existence of the thermodynamic limit of the Hopfield's free energy in the replica-symmetric scenario and, as corollaries, we recover the self-consistency equations for the related order parameters \cite{Amit}. In Sec. \ref{SezioneQuattro} we report the strategy of our interpolation along with all the technical details of the proofs of the theorems. Finally, in Sec. \ref{SezioneCinque}, we share our conclusions and outlook for future investigations.  

\paragraph{\red{State-of-the-art and novel results.}}
\red{In order to better contextualize the results here reported and highlight the novelties w.r.t. the existing literature, we give a summary of known rigorous results about the quenched free energy for the Hopfield model.
\begin{itemize}
    \item[{\it i)}] {\it The low-storage regime.} This is the well-understood regime of the Hopfield model. Several results are available in this case, concerning in particular convergence and self-averaging of the quenched free energy, description of phase transitions and symmetric (spurious) states \cite{Koch}, high-temperature perturbative expansions and its Borel summability \cite{Albeverio}, large deviation principles \cite{LDP}, the characterization of the Gibbs states \cite{Anton3} as well as scaling behaviors on the number of retrievable patterns \cite{mceliece1987capacity,newman1988memory,BovierBound}. In particular, Shcherbina and Tirozzi \cite{PST-lowstorage} proved that the free energy density of the model converges in probability to that of the Curie-Weiss model, the contribution of non-condensed patterns being negligible in this regime. Bovier and Gayrard \cite{BG} later established the almost sure convergence of the free energy to a deterministic limit, together with that of the measures induced on the overlaps. Besides the free energy, the fluctuations of the overlaps are also under control, both at the level of central limit theorems \cite{Gentz} and moderate deviations \cite{MD}.
    \item[{\it ii)}] {\it The high-storage regime.}  
      Although the picture is far less complete in the high-storage regime (the only unconditional result holding on the whole parameter space being self-averaging of the quenched intensive pressure \cite{PST-lowstorage,Anton3,Anton4}), rigorous results are still available. Among them, we stress Talagrand's proof \cite{Tala} on the validity of the replica-symmetric solution in a portion of the parameter space strictly containing the ergodic region. This was later refined with an analysis of the fluctuations of the order parameters, resulting in the convergence of their moments by means of exponential inequalities \cite{TalaNN1}. On the same footing, Bovier, Gayrard and Picco \cite{Anton2,Anton3,MS} characterized the Gibbs states and showed consistency with the replica-symmetric solution within a non-trivial, yet restricted, region of the phase diagram. Nevertheless, every unconditional result available in the literature is -- to date -- confined to a limited region of the parameter space. Allowing for concentration hypotheses on the order parameters was instead fruitful in confirming the replica-symmetric saddle point equations \cite{PST2}.  
    Such concentration-based conditional approach has now become a successful tool to derive the asymptotic quenched pressure in spin-glass models for machine retrieval within the Guerra's interpolation framework, see e.g. \cite{Fachechi1,pelatino,LindaRSB,rigorousdreaming}. As a further technical note, these results are typically derived by assuming the universality of quenched noise generated by non-retrieved patterns, adapting Carmona and Hu's \cite{Carmona} and Genovese's \cite{Genovese} arguments to the Hopfield model's setting.
\end{itemize}
Having established the state-of-the-art, we now move to mention the original results of the present manuscript:
\begin{itemize}
    \item [{\it a)}] {\it A novel, Guerra-like, interpolation working for neural networks.}  While, in the past, Guerra's interpolations played a pivotal role for understanding spin glasses, providing precious bounds to their free energies  (see e.g. \cite{guerra2001sum,Guerra1}), when applied to neural networks they yielded only weaker approximations (see e.g. \cite{barattolo,Guerra2}): in this manuscript, we introduce an interpolating strategy that leads to a well-posed variational problem, whose solution ultimately yields the replica-symmetric Hopfield free energy as a rigorous bound for the true one. Furthermore, this approach can now be adapted to a vast {\em zoology} of similar neural networks.
    \item [{\it b)}] {\it A sharp control of the thermodynamic limit, in the replica-symmetric regime, that returns the AGS picture.} In theoretical physics literature, replica symmetry is typically enforced by assuming self-averaging of the order parameters, i.e., by stating that their distributions are Dirac deltas centered on their expected values (see e.g. \cite{AGS1,PST1}).  Here, we relax this hypothesis and study how the concentration of these measures must occur in order to guarantee the replica-symmetric picture.
\end{itemize}
}

\section{Hopfield networks: generalities and related models}\label{SezioneDue}
In this Section we introduce the definitions and tools necessary to obtain our main results, presented in the next Section. 
\begin{Def} [Hopfield model] Let us consider a system of $N$ neurons, whose configuration is denoted by $\boldsymbol \sigma = (\sigma_1, \sigma_2, ..., \sigma_N) \in \Sigma_N := \{+1,-1\}^N$. Further, let us introduce a set of $K+1$ random vectors of length $N$, referred to as patterns, denoted as $\bb \eta \in \{-1, +1 \}^N$ and $\{\bb \xi ^{\mu} \}_{\mu=1,...,K} \in \mathbb R^{N\times K}$, such that $P(\eta_i=+1)=P(\eta_i=-1)=\frac{1}{2}$ for any $i \in \{1, ...,N\}$, and $\xi_i^\mu \underset{i.i.d.}\sim \mathcal{N}(0,1)$ for any $i \in \{1,...,N\}$ and $\mu \in \{1,...,K\}$. The Hamilton function of the Hopfield model is
	\begin{equation}\label{eq:HamiltonHopfield}
		H_{N,K,\bb\eta,\bb \xi}(\bb\sigma):=-\frac N2 m_{\bb\sigma}^2-\frac1{2N}{\sum_{i,j=1}^N}\sum_{\mu=1}^K \xi^\mu_i \xi^\mu_j  \sigma_i \sigma_j,
	\end{equation}
 where 
 \begin{equation}\label{zingarata}
 m_{\bb\sigma}:=\frac1N \sum_{i=1}^N \eta_i \sigma_i 
 \end{equation}
 is the Mattis magnetization of the pattern $\bb\eta$ to be retrieved.
\end{Def}
\begin{Rmk} 
\red{We assume that solely the pattern to be retrieved, i.e. $\boldsymbol{\eta}$ (also called ``condensed pattern'', see Def.~\ref{def:SPC}), displays binary entries, while all the not-retrieved ones, i.e. $\{\boldsymbol{\bb \xi^{\mu}}\}_{\mu=1,...,K}$, are equipped with  real entries:  there is no loss of generality in this choice as the quenched noise carried by non-condensed patterns is expected to exhibit the {\em universality property}\footnote{Universality refers to the fact that the expression of the quenched free energy in the thermodynamic limit does not depend on the fine details of the disorder distribution. In particular, the same limiting free energy is obtained whether the random variables $\xi_i^\mu$ are Rademacher distributed or standard Gaussian.} as proved by Carmona \& Hu \cite{Carmona} for the standard Sherrington-Kirkpatrick model and by Genovese \cite{Genovese} for bipartite spin glasses. For the Hopfield case, assuming this heterogeneity in the patterns yields the same AGS picture as shown in \cite{barattolo}; however, since this agreement alone does not constitute a general proof of the universality property, we provide a formalization thereof in App. \ref{app:universality}.}
\end{Rmk}
\begin{Def}\label{def:pressureH}
	Denoting with $\beta = T^{-1} \in \mR_+$ \red{with $T$ being} the fast noise (or {\em temperature} in physical jargon), the configuration space $\Sigma_N$ is endowed with the random Boltzmann-Gibbs measure 
\begin{equation}
    \label{eq:BoltzmannGibbs}
    \mP (\bb\sigma) = \frac{\exp(-\beta H_{N,K,\bb\eta,\bb \xi}(\bb\sigma))}{Z_{N,K,\bb\eta,\bb \xi,\beta}},
\end{equation}
with associated partition function
 \begin{equation}
     \label{eq:PartitionFunctionHopfield}
      Z_{N,K,\bb\eta,\bb \xi,\beta}:=\sum_{\bb\sigma \in \Sigma_N} \exp(-\beta H_{N,K,\bb\eta,\bb \xi}(\bb\sigma))
 \end{equation}
and intensive statistical pressure, at finite size $N$, defined as
\begin{equation}
\mathcal{A}_{N,K,\bb\xi,\bb \eta}(\beta):= \frac1N \log Z_{N,K,\bb\eta,\bb\xi,\beta}  
\end{equation}
while its  quenched expectation, at finite size $N$, reads 
\begin{equation}\label{eq:quenched_pressure_Hopfield}
A_{N,K}(\beta):= \frac1N\mE{\bb\xi,\bb \eta}\log Z_{N,K,\bb\eta,\bb\xi,\beta}  
\end{equation}
where $\mE{\bb\xi,\bb\eta}$ encompasses the Gaussian average w.r.t. the non-retrieved patterns $\bb\xi$ and the Rademacher average w.r.t. the retrieved pattern $\bb\eta$\footnote{The average w.r.t. $\bb\eta$ is actually redundant. In fact, as long as we are interested in the condensation of one single-pattern,  the quenched statistical pressure (and thus the quenched free energy) does not explicitly depend on the direction $\bb\eta$ as a direct consequence of the invariance of the Hamiltonian of the Hopfield model, see Eq. \eqref{eq:HamiltonHopfield}, w.r.t. the ``gauge'' transformation $\sigma_i\to\eta_i \sigma_i$.}. 
\end{Def}
\begin{Rmk}
In this paper, we do not deal with the intensive quenched {\em free energy} $F_{N,K}(\beta)$ (often preferred in the theoretical physical literature), rather we use the (intensive and quenched) {\em statistical pressure} $A_{N,K}(\beta)$ (often preferred in the mathematical physics literature): however, the two observables are trivially related by $A_{N,K}(\beta)=-\beta F_{N,K}(\beta)$ \cite{Guerra1,Fachechi1}.
\end{Rmk}
\begin{Rmk}
As proved by Bovier, Gayrard and Picco \cite{Anton3,Anton4},  in the thermodynamic limit the intensive statistical pressure of the Hopfield model is a self-averaging quantity around its quenched expression, namely $\lim_{N\to\infty} \left( \mathcal{A}_{N,K,\bb\xi,\bb \eta}(\beta) - A_{N,K}(\beta) \right)^2 = 0$, thus, in the following it is enough to focus on the quenched statistical pressure $A_{N,K}(\beta)$ (see Eq. \eqref{BovierMiss}) and prove its existence in the $N\to \infty$ limit.
\end{Rmk}
\par\medskip
As $N \to \infty$, we must specify how the number of patterns scales with the number of neurons,  separating the low-storage regime from the high-storage one, as specified by the next
\begin{Def}[high-storage and low-storage regimes] \label{def:TDL}
The load of the network at finite $N$ is introduced as $\alpha_N : = \frac{K(N)}{N}$, i.e. the ratio between the number  of patterns to store (i.e. $K$) and the number of neurons to handle them (i.e. $N$). Defining the asymptotic load $\alpha :=\lim_{N\to\infty}\alpha_N$, we say that the network operates in the high-load regime if $\alpha > 0$ or in the low-load regime if $\alpha=0$ and the limit $N\to\infty$ with $\alpha_N\to \alpha$ is referred to as the thermodynamic limit for the Hopfield model.
\end{Def}
\begin{Rmk}
In the low-storage regime ($\alpha=0$) the Hopfield model does not exhibit glassy behavior, which considerably simplifies its analysis. Indeed, the corresponding intensive quenched statistical pressure can be computed rigorously; see, for instance \cite{PST-lowstorage}.
\end{Rmk}
\begin{Rmk}    \label{rmk:alpha_N}
The behavior of the intensive quenched statistical pressure of the Hopfield model in the thermodynamic limit only depends on the asymptotic load $\alpha$ and not on the specific behavior of the sequence $K(N)$ at finite $N$, namely, if $K(N)$ and $K'(N)$ are two sequences with $\lim_N K(N)/N= \lim_N K'(N)/N=\alpha$, then their respective intensive statistical pressures become the same in the thermodynamic limit. 
\newline
Indeed, defining $\overline K(N) = \max(K(N),K'(N))$ and $\underline K(N) = \min(K(N),K'(N))$, we can interpolate between the Hopfield model with $\underline K(N)$ number of patterns and the one with $\overline K(N)$ with standard Guerra's approach. This way, one can show that
$$
\vert A_{N, K(N)}(\beta)- A_{N, K'(N)}(\beta)\vert=\vert A_{N,\overline K(N)}(\beta)- A_{N,\underline K(N)}(\beta)\vert \le \frac \beta2\frac {\overline K -\underline K}{\overline K} \mathbb E[\lambda_{max}(\bb J)],
$$
with $J_{ij}= \frac1N\sum_{\mu =1}^{\overline K}\xi^\mu_i \xi^\mu_j$ being the Hebb coupling matrix. Since $\overline K\sim \alpha N+ o(N)$ at large $N$ and $\alpha>0$, we have:
$$
\lim_{N\to\infty}\vert A_{N,\overline K}(\beta)- A_{N,\underline K}(\beta)\vert \le\frac \beta2\lim_{N\to\infty} \frac {\overline K -\underline K}{N} \frac{\mathbb E[\lambda_{max}(\bb J)]}\alpha=0,
$$
since $\mathbb E[\lambda_{max}(\bb J)]\to (1+\sqrt{\alpha})^2$, and $(\overline K-\underline K)/N\to 0$.
It immediately follows that $\lim_N \vert A_{N,K(N)}(\beta)-A_{N,K'(N)}(\beta)\vert=0$ for any $K(N)$, $K'(N)$ with the same scaling limit. 
From now on, for technical convenience but without loss of generality, we will work with the sequence $K(N)= \max(1,\lfloor{\alpha N}\rfloor)$ so that $K(N)\ge 1$ for all $N$.
\end{Rmk}
\begin{Def} The thermodynamic limit of the  intensive, quenched, statistical pressure for the Hopfield model in the high-storage regime is defined as
	\begin{equation}\label{BovierMiss}
		A(\alpha,\beta):= \lim_{N\to\infty} A_{N,K}(\beta).
	\end{equation}
\end{Def}

\begin{Def}[Boltzmann and quenched averages]
Given a function $O: \Sigma_N\to \mR$ of the neurons, the Boltzmann average is defined as
	\begin{equation}\label{eq:BoltzmannAverageHopfield}
		\Bav{O}{N,K,\bb\eta,\bb\xi,\beta}= Z_{N,K,\bb\eta,\bb \xi,\beta}^{-1}\sum_{\bb\sigma \in \Sigma_N}O(\bb\sigma)B_{N,K,\bb\eta,\bb\xi,\beta}(\bb\sigma),
	\end{equation}
	with $B_{N,K,\bb\eta,\bb\xi,\beta}(\bb\sigma)=\exp(-\beta H_{N,K,\bb\eta,\bb \xi}(\bb\sigma))$ being the Boltzmannfaktor. 
\newline 
The quenched average is then defined as 
\begin{equation}
\QBav{O}{N,K,\beta}=\mE{\bb\xi}\Bav{O}{N,K,\bb\eta,\bb\xi,\beta}.
\end{equation}
\end{Def}

\begin{Prp}[Equivalence of Hopfield neural networks and restricted Boltzmann machines]\label{Nabla}
The partition function \eqref{eq:PartitionFunctionHopfield} of the Hopfield model can be recast as
	\begin{es}
		\label{eq:Duality} 
		Z_{N,K,\bb\eta,\bb \xi,\beta}&:=\sum_{\bb\sigma \in \Sigma_N}\int _{\mR^K}d\mu_{K,\beta}(\bb z)\exp\Big(\frac{\beta N}{2}m_{\bb\sigma}^2+{\frac{\beta}{\sqrt{N}}}\sum_{i,\mu=1}^{N,K}\xi^\mu_i \sigma_i z_\mu\Big),
	\end{es}
where $d\mu_{K,\beta} (\bb z) = \prod_{\mu=1}^K (\frac{\beta}{2\pi})^{1/2}dz_\mu \exp(-\frac{\beta}2 z_\mu^2)$ is the $K$-dimensional Gaussian measure and $m_{\bb\sigma}$ is the Mattis magnetization defined in \eqref{zingarata}.
\end{Prp}

\begin{proof}
    The proof works by a straightforward application of the Hubbard-Stratonovich transformation on the $K$ Gaussian (non-retrieved) patterns.
\end{proof}
\begin{Rmk}
In the integral representation given by Prop. \ref{Nabla} we introduced $K$ hidden variables $z_{\mu} \in \mR,\ \mu \in \{1,...,K\}$ 
such that, in the exponential at the r.h.s. of Eq. \eqref{eq:Duality}, we can single out two contributions: the former corresponds to the Hamiltonian of a Mattis model (namely a simple Curie-Weiss-like model) and the latter corresponds to the Hamiltonian of a bipartite spin glass called {\em restricted Boltzmann machine} in Machine Learning jargon \cite{RBMoriginal}, that reads as
\begin{equation}\label{RBM-dual}
H_{N,K,\bb\eta,\bb \xi}(\bb\sigma, \bb z) := -\frac{1}{\sqrt{N}} \sum_{i,\mu=1}^{N,K}\xi^\mu_i \sigma_i z_\mu,
\end{equation}
where the $N$ Ising neurons $\boldsymbol{\sigma}$ belong to the {\em visible layer} while the $K$ Gaussian neurons $\boldsymbol{z}$ belong to the {\em hidden layer}. We called Gaussian neurons the hidden real-valued variables $\boldsymbol{z}$ because the measure $d\mu_{K,\beta} (\bb z)$ in \eqref{eq:Duality} can be interpreted as their Gaussian prior.\\
Given this duality, it is convenient to enlarge the configuration space to $\hat \Sigma_N = \Sigma_N \times \mR ^K$ so to host the $\boldsymbol z$ variables too. The Boltzmann-Gibbs measure for the Hopfield model is then represented by the marginal of the joint probability distribution (namely, the probability distribution $\hat \mP (\bb\sigma,\bb z)$ of the restricted Boltzmann machine) on $\hat \Sigma_N$, i.e. $\mP (\bb\sigma)= \hat \mP(\{\bb\sigma\}\times \mR^K)$. 
\end{Rmk}
%
\begin{Def}[Control parameters, order parameters and related fluctuations]  \label{CeOpm}
The statistical mechanical theory of the Hopfield model is worked out by describing how suitable macroscopic observables, referred to as {\em order parameters}, evolve in the space of the {\em control parameters}. The latter, in the thermodynamic limit, are $\alpha$ and $\beta$ while the former are
\begin{eqnarray}
m_a &\equiv& m_{\bb\sigma^{(a)}}:=\frac1N \sum_{i=1}^N \eta_i \sigma_i ^{(a)},\label{eq:m_def}\\
q_{ab} &\equiv& q_{\bb\sigma^{(a)},\bb\sigma^{(b)}}:=\frac1N \sum_{i=1}^N \sigma_i ^{(a)} \sigma_i ^{(b)},\label{eq:q_def}\\
p_{ab} &\equiv& p _{\bb z^{(a)},\bb z^{(b)}}:=\frac1K \sum_{\mu=1}^K z^{(a)}_\mu z^{(b)}_\mu\label{eq:p_def},
\end{eqnarray}
where the labels $(a)$ and $(b)$ are replica indices: $\{\bb \sigma^{(a)}, \bb z^{(a)} \}$ and $\{\bb \sigma^{(b)}, \bb z^{(b)}\}$ represent the configurations of two replicas $a, b$, that is, two realizations of the network at identical quenched noise. Further, once introduced $s \in \mN$, the $s$-replicated configuration space, where the overlaps are defined, $\hat \Sigma_N ^{(s)}:=\hat \Sigma_N ^{\otimes s}\equiv \Sigma_N^{\otimes s}\times \mR^{sK}$ is naturally endowed with the product Boltzmann-Gibbs measure $\mP_s=\bigtimes_{a=1}^s {\mP}^{(a)}$. For a function $O : \hat \Sigma_N^{(s)} \to \mathbb R$, this yields the $s$-replicated expectation $\langle O \rangle_{N,K,\beta}^{(s)}$. In particular, one has $\rQBav{m_a}{N,K,\beta}{1}$, $\rQBav{q_{ab}}{N,K,\beta}{2}$ and $\rQBav{p_{ab}}{N,K,\beta}{2}$ and their thermodynamic limit (see Def. \ref{def:TDL}) will be denoted by $\bar m = \lim_{N\to\infty} \rQBav{m_a}{N,K,\beta}{1}$, $\bar q = \lim_{N\to\infty} \rQBav{q_{ab}}{N,K,\beta}{2}$ and $\bar p = \lim_{N\to\infty} \rQBav{p_{ab}}{N,K,\beta}{2}$ for all $a,b=1,\dots,s$. 
\newline
The (suitably rescaled in $N$) fluctuations of the order parameters around their means are introduced as
\begin{eqnarray}\label{eq:mu-signal}
      \Delta_m \equiv \mu_a &:=& \sqrt N (m_a -\bar m),\\
      \Delta_q \equiv \theta_{ab} &:=& \sqrt N (q_{ab}-\bar q),\label{eq:theta_def}\\
	\sqrt{\alpha} \Delta_p \equiv \rho_{ab}&:=& \sqrt K (p_{ab}-\bar p). \label{eq:rho_def}
\end{eqnarray}
\end{Def}
To simplify the notation, we will omit the explicit dependence of the quenched averages on $K$ and $\beta$. 

As previously mentioned, we aim to establish a bridge between the Hopfield model and suitable mixtures of spin glasses. Specifically, we introduce a hard- and a soft-spin glass, as given by the next Definitions \ref{HSKdef} and \ref{SSKdef}, respectively, for which the existence of the asymptotic free energy has been previously established, as recalled in Theorems \ref{thm:GT} and \ref{GT-SSK}.

\begin{Def}[(Hard) Sherrington-Kirkpatrick model]\label{HSKdef}
    Let $\beta\in \mR_+$, $h\in \mR$ and $J_{ij}\underset{i.i.d.}\sim \mathcal N(0,1)$ for all $i,j =1,\dots,N$.    The Hamiltonian of the (hard) Sherrington-Kirkpatrick model \cite{sherrington1972solvable} is
        \begin{equation}
        \label{eq:SK-CF}
        H^{SK}_{N,\bb J, h}(\bb\sigma)= -\frac{1}{\sqrt{2N}}\sum_{i,j=1}^N J_{ij} \sigma_i \sigma_j -  h\sum_{i=1}^N \sigma_i,
    \end{equation}
    where the $N$ binary variables are $\sigma_i = \pm 1,\ i \in \{1,...,N\}$.
    \newline
    The partition function of the Sherrington-Kirkpatrick model is defined as
    \begin{equation}
        \label{eq:SKPartition}
        Z^{SK}_{N,\bb J,\beta,h}=\sum_{\bb\sigma \in \Sigma_N}\exp\Big(\frac{\beta}{\sqrt{2N}}\sum_{i,j=1}^N J_{ij} \sigma_i \sigma_j + \beta h\sum_{i=1}^N \sigma_i\Big).
    \end{equation}
    The intensive quenched statistical pressure at finite volume of the Sherrington-Kirkpatrick model \eqref{eq:SK-CF} is defined as 
\begin{equation}    
    A_N^{SK}(\beta,h)=\frac1N \mE{\bb J} \log Z^{SK}_{N,\bb J,\beta,h},
\end{equation}
where $\mE{\bb J}$ denotes the expectation with respect to $\boldsymbol J$ and $A^{SK}(\beta,h) = \lim_{N\to \infty} A_N^{SK}(\beta,h)$.
\end{Def} 

\begin{Thm}[thermodynamic limit, hard SK: Guerra-Toninelli]\label{thm:GT}
	The thermodynamic limit of the intensive quenched pressure for the Sherrington-Kirkpatrick model \eqref{eq:SK-CF} exists, and it is equal to
	\begin{equation}
		\label{eq:GuerraToninelli}
		A^{SK}(\beta,h) = \lim_{N\to\infty}A_{N }^{SK}(\beta,h)= \underset{N}{\mathrm{sup}}\, A_{N }^{SK}(\beta,h).
	\end{equation}
\end{Thm}
The result is reported in the celebrated paper \cite{guerra2002thermodynamic} and it is based on the super-additivity property of the intensive pressure, together with Jensen inequality and Fekete lemma \cite{fekete1923verteilung}. 
\par\medskip

\begin{Def}[(Soft) Sherrington-Kirkpatrick model]\label{SSKdef}
Let $\beta,\lambda\in \mR_+$ and $\bar{J}_{\mu\nu}\underset{i.i.d.}\sim\mathcal N(0,1)$ for all $\mu,\nu =1,\dots,K$. The Hamiltonian of the (soft) Sherrington-Kirkpatrick model is
\begin{equation}
\label{eq:SSK-CF}
H^{sSK}_{K,\bb J}(\bb z)= -\frac{1}{\sqrt{2K}}\sum_{\mu,\nu=1}^K  \bar{J}_{\mu\nu}z_\mu z_\nu,
\end{equation}
where the $K$ real variables $z_{\mu} \in \mR,\  \mu \in \{1,...,K\}$.
    \newline 
The partition function of the (soft) Sherrington-Kirkpatrick model is defined as
	\begin{equation}
		\label{eq:PartitionFunctionSoftSK}
		Z_{K,\bb{\bar J},\beta,\lambda }^{sSK}= \int_{\mR^K} dP_{\beta,\lambda}(\bb z)\exp\Big(\frac \beta{\sqrt {2K}}\sum_{\mu,\nu=1}^K  \bar{J}_{\mu\nu}z_\mu z_\nu\Big),
	\end{equation}
with 
$$dP_{\beta,\lambda}(\bb z)= d\mu_K (\bb z)\exp\big( -\frac {\beta^2}{4K}\big(\sum_{\mu=1}^K z_\mu^2\big)^2+\frac{\lambda}2 \sum_{\mu=1}^K z_\mu^2\big),$$ 
and $d\mu_{K} (\bb z) = \prod_{\mu=1}^K ({2\pi})^{-1/2}dz_\mu \exp(-\frac{1}2 z_\mu^2)$ being the standard Gaussian measure in $K$-dimension. 
\newline
The  intensive, quenched, statistical pressure, of the (soft) Sherrington-Kirkpatrick model at finite volume $K$, is defined as 
\begin{equation}
A_{K}^{sSK}(\beta,\lambda)= \frac1K \mE{\bb {\bar J}} \log Z_{K ,\bb {\bar J},\beta,\lambda}^{sSK},
\end{equation}
where $\mE{\bb {\bar J}}$ denotes the expectation with respect to $\bb{\bar J}$  and $A^{sSK}(\beta,\lambda)=\lim_{K \to \infty} A_{K}^{sSK}(\beta,\lambda)$.
\end{Def}

\begin{Rmk}\label{rmk:limit_beta}
The intensive quenched statistical pressure of the soft Sherrington-Kirkpatrick model is uniformly bounded by
\begin{es}
    A_K^{sSK}(\beta,\lambda)\le A (\beta,\lambda,p)= \log \sigma_{\beta,\lambda}(p) +\frac{\beta^2 p \sigma_{\beta,\lambda}(p)^2}{2}+\frac{\beta^2 p^2}{4},
\end{es}
for any $p\in \mR_+$ and $\sigma_{\beta,\lambda}(p) = (1-\lambda+\beta^2 p)^{-1/2}.$ This holds in particular for $p$ minimizing the upper bound of the inequality, namely $p=(\beta-(1-\lambda))/\beta^2$ if $\lambda >1$ or $\lambda<1$ and $\beta>1-\lambda$, and $p=0$ otherwise. In the former case, we would thus have $A_K^{sSK}(\beta,\lambda)\le-\frac 12 \log \beta +(1-\lambda-\beta)(1-\lambda-3\beta)/(4\beta^2)$ uniformly in $K$. This implies that, regardless of the value of $\lambda$, $\lim_{\beta\to\infty} A_K ^{sSK} (\beta,\lambda)=-\infty$ for all $K \in \mN$ if $p>0$.\footnote{
The parameter $\beta$ controls the radius $R_*(\beta,\lambda)$ of the dominant contributions to the partition function. In particular, $R_*$ decreases as $\beta$ is increased. In the limit $K\to\infty$, where fluctuations of the quenched pressure vanish, the integral concentrates at $R_*=0$ (a degenerate sphere). Consequently, the free energy diverges to $-\infty$ as $\beta\to +\infty$.
}
\end{Rmk}

\begin{Thm}[thermodynamic limit, soft SK: Barra-Genovese-Guerra-Tantari]\label{GT-SSK}
 The thermodynamic limit of the intensive quenched statistical pressure for the soft Sherrington-Kirkpatrick model \eqref{eq:PartitionFunctionSoftSK} exists and it is equal to
 \begin{equation}
 	\label{eq:BGGT}
 		A^{sSK}(\beta,\lambda) = \lim_{K\to\infty}A_{K}^{sSK}(\beta,\lambda)= \underset{K}{\mathrm{sup}}\, A_{K}^{sSK}(\beta,\lambda).
 \end{equation}
\end{Thm}
The proof is reported in \cite{barra2014solvable} and  it is based again on the super-additivity
property of the intensive pressure, together with Jensen inequality and Fekete lemma.

\section{Main results: the replica-symmetric scenario and AGS theory}\label{SezioneTre}
The related proofs are detailed in the next Section \ref{SezioneQuattro}. 

\begin{Rmk}
In our proofs we interpolate the intensive quenched statistical pressure of the Hopfield model with that of a mixture of a hard and a soft spin glass and call $t \in [0,1]$ the interpolation parameter. As a consequence, we introduce a deformed Boltzmann measure \cite{Aizenman} and all the averages taken under this deformed measure will be denoted by a subscript $t$, i.e. $\langle . \rangle_N \to \langle . \rangle_{N,t}$, so that quenched expectation values of observables (as well as the corresponding fluctuations) will be functions of the interpolating parameter $t$. The deformed Boltzmann measure also depends on the auxiliary variables $(\psi, \beta_1) \in \mathbb R \times \mathbb R_+$, see Def.~\ref{InterMilan}. In particular, the thermodynamic limit of quenched averages of the order parameters will be denoted as $\bar m_t(\psi, \beta_1)$, $\bar q_t(\psi, \beta_1)$ and $\bar p_t(\psi, \beta_1)$.
\end{Rmk}
Given this premise, we can now introduce the next
\begin{Def}[Self-averaging property]\label{def:weak_sap}
An observable $O: \hat \Sigma_N ^{(s)}\to \mR$ is said to satisfy the self-averaging property (SAP) at order $r\in \mN$ if: 
	\begin{itemize}
		\item[i)]   the quenched expectation value $\rQBav{O}{N,t}{s}$ converges point-wise to a (bounded in $t\in [0,1]$) Lebesgue-measurable limit $\bar O_t$ as $N\to\infty$;
        \item[ii)] it has bounded moments up to order $r$, namely $\sup_{N\in \mN} \vert\rQBav{O^n}{N,t}{s}\vert <\infty$ for $n=1,\dots,r$ and all $t\in [0,1]$;
		\item[iii)] the rescaled fluctuation $\Delta_O= \sqrt N ( O-\bar O_t)$ satisfies $\underset{t\in[0,1]}\esssup\,\underset{{N\in \mN}}\sup\langle \vert \Delta_O \vert ^r \rangle_{N,t} ^{(s)}:= M^{(r)}_O <\infty$.
	\end{itemize}
Furthermore, a family of observables $\{O_l\}_{l=1}^L$ satisfies SAP at the orders $(r_l)_{l=1}^L$ if any element $O_l$ satisfies SAP at the order $r_l$.
\end{Def}
Since any observable $O$ is generally defined on $s$ replicas (for instance, the overlaps $q_{ab}$ are functions of two replicas), the quenched averages $\rQBav{O}{N,t}{s}$ are unambiguously defined. However, to keep the notation self-consistent, we also dropped the dependence on $s$ of the thermodynamic limit $\bar O_t$ (indeed, the crucial dependence in this setup concerns the interpolating parameter).

\begin{Rmk}
    The SAP implies some analytical and probabilistic properties, as reported hereafter.
    \begin{itemize}
    \item
Assume that an observable $O$ satisfies SAP at order 1. By $iii)$, there exists a dense set $D\subset [0,1]$ for which $\sup_{t\in D} \vert \rQBav{O}{N,t}{s}-\bar O_t\vert\to 0$, thus $ \rQBav{O}{N,t}{s}$ converges uniformly in $D$. Also, since for all $N\in \mN$ the function $ \rQBav{O}{N,t}{s}$ is continuous, $\rQBav{O}{N,t}{s}$ converges uniformly to $\bar O_t$  for all $t\in [0,1]$, and the point-wise limit $\bar O_t$ is $t$-continuous on $[0,1]$.
\item
Assume that an observable $O$ satisfies SAP at order $r$. By $iii)$ and by Chebyshev inequality we have
$\mP_t (\vert \Delta_O \vert \ge x)\le x^{-r}{\langle \vert \Delta_O \vert ^r \rangle_{N,t}}$. 
The tail behavior of the probability distribution of the fluctuation is therefore stable under the thermodynamic limit, as
$$
 \lim_{N\to \infty}\mP_{t,N} (\vert \Delta_O \vert \ge x)\le \sup_{N\in \mN}\mP_{t,N }(\vert \Delta_O \vert \ge x)\le \sup_{N\in \mN}\frac{\langle \vert \Delta_O \vert ^r \rangle_{N,t}}{x^r}\le \frac{M_O ^{(r)}}{x^r},
$$
for almost any $t\in [0,1]$, thus implying that there is no probability mass escape at infinity in the thermodynamic limit, (almost) regardless of the value of $t$.
Further, the moments $\langle \vert \Delta_O \vert ^n\rangle_t$ are almost everywhere finite in the thermodynamic limit also for any $n=1,\dots,r-1$ because of Jensen's inequality: $\langle\vert \Delta_O \vert ^n\rangle_{N,t} ^{(s)}\le \big[\langle\vert \Delta_O \vert ^r\rangle_{N,t} ^{(s)}\big]^{\frac nr}$ for $n=1,\dots,r-1$. In particular $\lim_{N\to\infty}\langle\vert \Delta_O \vert ^n\rangle_{N,t} ^{(s)}\le \lim_{N\to\infty}\big[{\langle\vert \Delta_O \vert ^r\rangle_{N,t} ^{(s)}}\big]^{\frac n r}\le (M_{O}^{(r)})^{\frac nr}:= M^{(n)}_{O}$ for almost any $t\in[0,1]$.
 \end{itemize}  
\end{Rmk}

\begin{Rmk}\label{rmk:VIP}
    If an observable $O$ of $s$ replicas satisfies SAP (at least) at order 2, then trivially $\lim_{N\to\infty} (\rQBav{O^2}{N,t}{s}-(\rQBav{O}{N,t}{s})^2)=0$. If $F$ is a function of $r$ replicas with bounded second moment, then by Cauchy-Schwarz inequality
    $$
    \lim_{N\to \infty}[\rQBav{O F}{N,t}{\nu}-\rQBav{O}{N,t}{s}\rQBav{F}{N,t}{r}]^2\le\lim_{N\to \infty}[\rQBav{(O-\rQBav{O}{N,t}{s})^2 }{N,t}{s}\cdot \rQBav{F^2}{N,t}{r}]=0,
    $$
    with $\nu = \max\{r,s\}$. Exploiting SAP, it is in particular possible to show that the error in the factorization is of order $N^{-1/2}$, as $\sup_N \sqrt N \vert\rQBav{O F}{N,t}{\nu}-\rQBav{O}{N,t}{s}\rQBav{F}{N,t}{r}\vert\le C$ for some $C>0$.
\end{Rmk}

We are now ready to state the main result of this paper:

\begin{Thm}
     [Existence of thermodynamic limit (RS)]\label{thm:thermolim}
     Let $(\alpha,\beta)\in \mR_+^2$ and let $K=K(N)$ such that $\lim_N K(N)/N=\alpha$. 
     The thermodynamic limit $A(\alpha,\beta)$ of the intensive quenched statistical pressure of the Hopfield model $A_{N,K}(\beta)$ exists and, in particular,
   \begin{equation}\label{eq:main_thm_RS}
        A(\alpha,\beta) = \max_{\psi\in \mR}\max_{\beta_1>0}\Big[A^{SK}\big(\beta_1,\frac{\beta \psi}{\beta_1}\big)+\alpha A^{sSK}\big(\frac{\sqrt \alpha\beta}{\beta_1},\beta\big)-\frac{\beta \psi^2}{2}-\frac{\beta_1 ^2}{4}\Big],
    \end{equation}
    as long as the family $(m_1,q_{12},p_{12})$ satisfies SAP at order $4$ for the values
   $(\psi^*, \beta_1^*)$ which solve the implicit equations $\psi=\bar m_{t=0}(\psi,\beta_1)$ and $\beta_1^2\bar q_{t=0}(\psi,\beta_1)=\alpha\beta \bar p_{t=0}(\psi,\beta_1)$. 
\end{Thm}

 
\begin{Cor}[Amit-Gutfreund-Sompolinsky formula]\label{thm:AGS}
The solution of the variational problem formulated in Eq. \eqref{eq:main_thm_RS} is given by the AGS formula, namely
\begin{es} \label{eq:32}
    A(\alpha,\beta)&= -\frac{\beta \psi^2}{2}+\int _\mR d\mu(\tau)\log\left(2\cosh(\beta \psi+\tau\sqrt{\alpha \beta p})\right)-\frac{\alpha\beta}{2}p (1-q)\\&+\frac\alpha 2 \frac{ \beta q}{1-\beta(1-q)}-\frac\alpha2 \log[1-\beta(1-q)],
\end{es}
with $\psi,p,q$ satisfying the self-consistency equations
\begin{eqnarray}
    \psi &=& \int _\mR d\mu (\tau) \tanh (\beta \psi+ \tau\sqrt {\alpha\beta p}),\\
    q &=& \int _\mR d\mu (\tau) \tanh ^2(\beta \psi+ \tau\sqrt {\alpha\beta p}),\\
    p &=&\frac{\beta q}{[1-\beta(1-q)]^2}.
\end{eqnarray}
\end{Cor}
The result in \eqref{eq:32} follows naturally from the concentration of measure of the order parameter. 

\section{Interpolation strategy and proof of the main theorem}\label{SezioneQuattro}
In this Section we provide all the arguments that we need to achieve the conclusions shown in Theorem \ref{thm:thermolim} and its related Corollary \ref{thm:AGS}. In particular in Sec. \ref{4.1} we construct an interpolating scheme that bridges the original Hopfield model with a proper mixture of a hard and a soft spin glass: we obtain a sum rule with a controllable remainder that is a function of the order parameters' fluctuations on the deformed measure depending on the interpolation parameter and that has a definite sign. Next, in Sec. \ref{4.2}, we explore the effect of the interpolation on the behavior of the order parameters and of their fluctuations and then, in Sec. \ref{4.3}, we use this control to bound the remainder term stemming from the sum rule, so as to reach our conclusions in the original (i.e., unperturbed by the interpolation) Boltzmann measure. Finally, in Sec. \ref{4.4} we set up our variational principle and achieve the theses of Theorem \ref{thm:thermolim} and Corollary \ref{thm:AGS}.

\subsection{Sum rules in the adapted Guerra interpolating framework}\label{4.1}  
After having introduced the involved models (i.e., the Hopfield Hamiltonian  \eqref{eq:HamiltonHopfield}, its dual representation provided by the restricted Boltzmann machine \eqref{RBM-dual}, the hard \eqref{eq:SK-CF} and the soft \eqref{eq:SSK-CF} spin glasses), we can introduce our interpolation strategy by the next
\begin{Def}\label{InterMilan}
	Let $t\in [0,1]$, $\beta_1,\beta_2 \in \mR_+$, $\psi\in \mR$, $J_{ij}, \bar J_{\mu\nu}\underset{i.i.d}\sim \mathcal N(0,1)$ for all $i, j= \{1,\dots,N\}$ and $\mu ,\nu = \{1,\dots,K\}$. We introduce the interpolating partition function
 \begin{es}
     \label{eq:InterpolatingPart}
     &Z_{N,K,\bb\eta,\bb\xi, \bb J ,\bb{\bar J},\bb \beta,\psi,t}=
     \sum_{\bb\sigma\in \Sigma_N}\int _{\mR^K}d\mu_K (\bb z)\exp\Big( t \frac{\beta N}{2} m_{\bb\sigma} ^2+\sqrt t \sqrt{\frac\beta N}\sum_{i,\mu=1}^{N,K}\xi^\mu_i \sigma_i z_\mu +(1-t)N \beta \psi m_{\bb\sigma}\\
		&+\sqrt {1-t}\big(\beta_1 \sqrt{\frac N2}\mc K_{\bb J}(\bb\sigma)+\beta_2 \sqrt{\frac K2}\mc G_{\bb {\bar J}}(\bb z)\big)+\frac
		{1-t}2 \big(   K\beta p_{\bb  z,\bb z}-\beta_2 ^2\frac K2 p_{\bb z,\bb z}^2\big)-\frac{1-t}4 N \beta_ 1^2 -\frac{1-t}{2}N \beta \psi^2\Big),
 \end{es}
 with $\bb \beta = (\beta,\beta_1,\beta_2)$, $d\mu_{K} (\bb z) = \prod_{\mu=1}^K ({2\pi})^{-1/2}dz_\mu \exp(-\frac{1}2 z_\mu^2)$ being the standard Gaussian measure in $K$-dimension, and
\begin{equation}\label{eq:QuadraticForms}
	\mc K_{\bb J}(\bb\sigma) =  \frac1N \sum_{i,j=1}^N J_{ij}\sigma_i \sigma_j,\quad
	\mc G_{\bb {\bar J}}(\bb z)  = \frac1K \sum_{\mu,\nu=1}^K \bar J_{\mu\nu}z_\mu z_\nu.
\end{equation}
The interpolating statistical pressure $\varphi_{N,K,\bb\beta,\psi,t}$ is defined as
\begin{es}\label{eq:InterpolatingPress}
    \varphi_{N,K,\bb\beta,\psi,t}= \frac1N \mE{\bb \xi,\bb J,\bb{\bar J}}\log Z_{N,K,\bb\eta,\bb\xi, \bb J ,\bb{\bar J},\bb \beta,\psi,t}.
\end{es}
\end{Def}

\begin{Def}\label{def:RS_all_s_replicated}
    Let $s \in \mN$, the $s$-replicated configuration space $\hat \Sigma_N ^{(s)}:=\hat \Sigma_N ^{\otimes s}\equiv \Sigma_N^{\otimes s}\times \mR^{sK}$ is naturally endowed with the product Boltzmann-Gibbs measure $\mP_s=\bigtimes_{a=1}^s {\mP}^{(a)}$ with associated partition function
	\begin{equation}\label{eq:ReplicatedPartitionFunctionHopfield}
	Z_{N,K,\bb\eta,\bb\xi, \bb J ,\bb{\bar J},\bb \beta,\psi,t}^{(s)}:= Z_{N,K,\bb\eta,\bb\xi, \bb J ,\bb{\bar J},\bb \beta,\psi,t}^s =\sum_{\bb\sigma_s \in \Sigma_N^{\otimes s}}\int _{\mR^{sK}}d\mu_{sK} (\bb z_s)\prod_{a=1}^s B_{N,K,\bb\eta,\bb\xi, \bb J ,\bb{\bar J},\bb \beta,\psi,t}(\bb \sigma^{(a)},\bb z^{(a)}),
\end{equation}
with $\bb\sigma_s =(\bb\sigma^{(1)},\dots,\bb \sigma^{(s)})$, $\bb z_s =(\bb z ^{(1)},\dots,\bb z^{(s)})$, and $B_{N,K,\bb\eta,\bb\xi, \bb J ,\bb{\bar J},\bb \beta,\psi,t}(\bb \sigma,\bb z)$ being the Boltzmannfaktor appearing in the interpolating partition function \eqref{eq:InterpolatingPart}. Given a function $O:\hat\Sigma_N^{(s)}\to \mR$ of the $s$-replicated system, the Boltzmann average is
\begin{equation}\label{eq:ReplicatedBoltzmannAveragenHopfield}
	\rBav{O}{N,K,\bb\eta,\bb\xi, \bb J ,\bb{\bar J},\bb \beta,\psi,t}{s}= \frac {\sum_{\bb\sigma_s \in \Sigma_N^{\otimes s}}\int _{\mR^{sK}}d\mu_{sK} (\bb z_s)O(\bb\sigma_s,\bb z_s)\prod_{a=1}^s B_{N,K,\bb\eta,\bb\xi, \bb J ,\bb{\bar J},\bb \beta,\psi,t}(\bb\sigma^{(a)},\bb z^{(a)})}{Z_{N,K,\bb\eta,\bb\xi, \bb J ,\bb{\bar J},\bb \beta,\psi,t}^{(s)}}.
\end{equation}
The quenched average is defined as $\rQBav{O}{N,K,\bb\beta,\psi,t}{s}= \mE{\bb\xi,\bb J,\bb{\bar J}}\rBav{O}{N,K,\bb\eta,\bb\xi, \bb J ,\bb{\bar J},\bb \beta,\psi,t}{s}.$
\end{Def} 
Note that, for the sake of clarity, we dropped the dependence on $\bb\eta$ in the quenched averages. We will also omit the dependence on $K$ (recalling that the storage capacity $\alpha_N=K/N$ relates the two volumes $K$ and $N$) and on $\bb \beta$, as the focus here is on the behavior of the intensive quenched statistical pressure as a function of $N$ and $t$.

\begin{Prp}\label{lem:mainprop}
    Setting $\beta_1 \beta_2 = \sqrt{\alpha}\beta$, the intensive quenched statistical pressure of the Hopfield model \eqref{eq:quenched_pressure_Hopfield} admits the following decomposition in terms of a hard and a soft spin glass
    \begin{equation}
    \begin{split}
        \label{eq:HopfieldDecomposition}
        A_{N,K}(\beta)&= A_N ^{SK}\big(\beta_1,\frac{\beta\psi}{\beta_1}\big)+\alpha_N A_K ^{sSK}(\beta_2,\beta)-\frac{\beta_1 ^2}{4}-\frac{\beta \psi^2}{2}\\&+\int _0 ^1 dt\, \Big(\frac1{2N}\rQBav{\mc Q_1}{N,t}1+\frac{\alpha_N}{2\alpha N}\rQBav{ \mc R_{12}}{N,t}2+\frac{(\alpha-\alpha_N)\beta_1 ^2 }{4\alpha}\rQBav{q_{12}^2}{N,t}{2}\Big),      
    \end{split}
    \end{equation}
    with $\mc Q_a = \beta N(m_a-\psi)^2$ and $\mc R_{ab}= \frac N2(\beta_1 q_{ab}-\sqrt{\alpha}\beta_2 p_{ab})^2$. In particular, if $K(N)=\max(1,\lfloor \alpha N \rfloor)$, for all $N>\alpha^{-1}$
    \begin{equation}
    A_{N,K}(\beta)\ge A_N ^{SK} \big(\beta_1,\frac{\beta\psi}{\beta_1}\big)+\alpha_N  A_K^{sSK} (\beta_2,\beta) -\frac{\beta_1 ^2}{4}-\frac{\beta \psi^2}{2}.
        \label{eq:HopfieldDecompositionIneq}
    \end{equation}
\end{Prp}
\begin{proof}
Let us consider the intensive quenched statistical pressure associated to the interpolating partition function \eqref{eq:InterpolatingPart}:
\begin{es}
&    \vphi_{N} (t)=\frac1N \mE{\bb\xi,\bb J,\bb{\bar J}}\log \sum_{\bb\sigma\in \Sigma_N}\int _{\mR^K}d\mu_K (\bb z)\exp\Big( t \frac{\beta N}{2} m_{\bb\sigma} ^2+\sqrt t \sqrt{\frac\beta N}\sum_{i,\mu=1}^{N,K}\xi^\mu_i \sigma_i z_\mu +(1-t)N \beta \psi m_{\bb\sigma} \\
		&-\frac{1-t}{2}N \beta \psi^2+\sqrt {1-t}\big(\beta_1 \sqrt{\frac N2}\mc K_{\bb J}(\bb\sigma)+\beta_2 \sqrt{\frac K2}\mc G_{\bb {\bar J}}(\bb z)\big)+\frac
		{1-t}2 \big(   K\beta p_{\bb  z,\bb z}-\beta_2 ^2\frac K2 p_{\bb z,\bb z}^2\big)-\frac{1-t}4 N \beta_ 1^2\Big).\nonumber
\end{es}
At $t=1$, it exactly gives the intensive quenched statistical pressure of the Hopfield model, i.e.
	\begin{es}\label{eq:phi1}
		\vphi _N (1) &=\frac1N \mE{\bb\xi}\log \sum_{\bb\sigma \in \Sigma_N}\int _{\mR^K}d\mu_K (\bb z)\exp\Big(\frac{\beta N}{2}m_{\bb\sigma}^2+\sqrt{\frac\beta N}\sum_{i,\mu=1}^{N,K}\xi^\mu_i \sigma_i z_\mu\Big)=\\
  &=\frac 1N \mE{\bb \xi}\log \sum_{\bb\sigma\in \Sigma_N}\exp\Big(\frac{\beta N}{2}m_{\bb\sigma}^2 + \frac{\beta}{2N}\sum_{i,\mu=1}^{N,K} \xi^\mu_i \xi^\mu_j \sigma_i \sigma_j\Big) = A_{N,K}(\beta).
	\end{es}
On the other hand, at $t=0$, we have
\begin{es}\label{eq:t=0}
	\vphi _N (0)&= \frac1N \mE{\bb J} \log \sum_{\bb\sigma\in \Sigma_N}\exp\big(N \beta \psi m_{\bb\sigma}+\beta_1 \sqrt{\frac N2}\mc K _{\bb J}(\bb\sigma)\big)\\&+\frac1N \mE{\bb {\bar J}} \log \sum_{\bb\sigma\in \Sigma_N}\exp\big(\beta_2 \sqrt{\frac K2}\mc G_{\bb {\bar J}} (\bb z) +\frac{K \beta}2 p_{\bb z,\bb z}-\frac {\beta_2^2 K }4 p_{\bb z,\bb z}^2\big) -\frac{\beta \psi^2}{2}-\frac{\beta_1 ^2}{4}.
\end{es}
Focusing on the first term in the r.h.s. in the above expression, we see that
\begin{equation*}
     \quad \frac1N \mE{\bb J} \log \sum_{\bb\sigma\in \Sigma_N}\exp\big(N \beta \psi m_{\bb\sigma}+\beta_1 \sqrt{\frac N2}\mc K _{\bb J}(\bb\sigma)\big)=\frac1N \mE{\bb J} \log \sum_{\bb\sigma\in \Sigma_N}\exp\big( {\frac{ \beta_1}{\sqrt{ 2N}}}\sum_{i,j=1}^N J_{ij}\sigma_i\sigma_j+{\beta\psi}  \sum_{i=1}^N \eta_i \sigma_i\big).
\end{equation*}
Using the $\mathbb Z_2$-invariance of the model, i.e. $\sigma_i\to \sigma_i\eta_i$, and  the fact that $J_{ij}\eta_i \eta_j$ is still $\mc N(0,1)$-distributed, we recognize
\begin{equation}\label{eq:phi0}
    \vphi _N (0)=A_{N}^{SK}\big(\beta_1,\frac{\beta \psi}{\beta_1}\big)+\alpha _NA_{K}^{sSK}(\beta_2,\beta)-\frac14 \beta_1 ^2-\frac{\beta \psi^2}{2}.
\end{equation}
These two extrema of the interpolative scaffold can be connected by the fundamental theorem of calculus, 
\begin{equation}\label{eq:GuerraPropagation}
	\vphi_N (1)= \vphi_N (0)+\int _0 ^1 dt \,\partial_{t} \vphi_N (t).
\end{equation}
To exploit this sum rule we need to compute the $t$-derivative of the interpolating quenched statistical pressure. We write
\begin{es}
	\label{eq:computation1}
	\partial_t \vphi_N (t)&=
 \frac1N \mE{\bb\xi,\bb J,\bb{\bar J}}\Big(\frac{\beta N}2 \omega_{N,t}(m_{\bb\sigma}^2) +\frac1{2\sqrt t} \sqrt{\frac{\beta}N}\sum_{i\mu}\xi^\mu_i \Bav{\sigma_i z_\mu}{N,t}-\beta N \psi \omega_{N,t}(m_1)+\frac{\beta N}{2}\psi^2\\&
 -\frac{1}{2\sqrt{1-t}}\frac{\beta_1}{\sqrt {2N}}\sum_{ij}J_{ij}\Bav{\sigma_i\sigma_j}{N,t}
	-\frac{1}{2\sqrt{1-t}}\frac {\beta_2}{\sqrt{2 K}}\sum_{\mu\nu}\bar J_{\mu\nu}\Bav{z_\mu z_\nu}{N,t}-\frac \beta 2\sum_\mu\Bav{z_\mu^2}{N,t}\\&
 +\frac{\beta_2^2}{4K}\sum_{\mu,\nu} \Bav{z_\mu^2z_\nu^2}{N,t}+\frac N4 \beta_1^2\Big).
\end{es}
The disorder-dependent contributions can be further simplified by using the Wick-Isserlis lemma. For instance, by using the definition of the order parameters provided in Eqs. \eqref{eq:q_def}-\eqref{eq:p_def}, for the second contribution we have
\begin{es}
	\label{eq:firstContribution}
	\sum_{i,\mu}\mE{\bb \xi}\xi^\mu_i \Bav{\sigma_i z_\mu}{N,t}&=\sum_{i,\mu} \mE{\bb \xi}\partial_{\xi^\mu_i} \Bav{\sigma_i z_\mu}{N,t}=\sqrt t \sqrt{\frac{\beta}N }\sum_{i\mu}\mE{\bb\xi}(\Bav{\sigma_i ^ 2z _\mu^2}{N,t}-\Bav{\sigma_iz _\mu}{N,t}^2)=\\
	&=\sqrt t \sqrt{\frac \beta N}N K (\rQBav{p_{11}}{N,t}2-\rQBav{q_{12 }p_{12}}{N,t}2).
\end{es}
The other contributions can be evaluated similarly. At the end of the computations, we are left with
\begin{es}
	\label{eq:computation2}
	\partial_t \vphi_N (t)&= 
    \frac{\beta}{2}\rQBav{(m_1-\psi)^2}{N,t}{1}-\frac{\alpha_N \beta }{2} 
	\rQBav{q_{12}p_{12}}{N,t}2
	+\frac{\beta_1^2 }{4} \rQBav{q_{12}^2}{N,t}2
	+\frac{\alpha_N\beta_2 ^2 }{4 }\rQBav{p_{12}^2}{N,t}2.
\end{es}
We now add and subtract the quantity $-\frac{\alpha \beta }{2} \rQBav{q_{12}p_{12}}{N,t}2+\frac{\alpha\beta_2 ^2 }{4 }\rQBav{p_{12}^2}{N,t}2$, such that we can rewrite the derivative as
\begin{es}
	\label{eq:computation2.1}
	\partial_t \vphi_N (t)&=
\frac{\beta}{2}\rQBav{(m_1-\psi)^2}{N,t}{1}
 +\frac{1}{4}(\beta_1 ^2\rQBav{q_{12}^2}{N,t}{2}-2\alpha \beta \rQBav{q_{12}p_{12}}{N,t}{2}+\alpha \beta_2 ^2 \rQBav{p_{12}^2}{N,t}{2})
\\&- \frac{(\alpha-\alpha_N)  }{4\alpha} (\beta_1^2 \rQBav{q_{12}^2}{N,t}{2}-2\alpha\beta \rQBav{q_{12}p_{12}}{N,t}{2}+\alpha \beta _2 ^2\rQBav{p_{12}^2}{N,t}2 )+\frac{(\alpha-\alpha_N)\beta_1 ^2 }{4\alpha}\rQBav{q_{12}^2}{N,t}{2}.
\end{es}
Choosing the tunable temperatures of the spin-glass models such that $\beta_1 \beta_2 =\sqrt {\alpha} \beta$, the contributions in round brackets form perfect squares, so that
\begin{es}
	\label{eq:computation3}
	\partial_t \vphi_N (t)&= 
\frac{\beta}{2}\rQBav{(m_1-\psi)^2}{N,t}{1}
 +\frac{\alpha_N}{4\alpha }\rQBav{(\beta_1 q_{12}-\sqrt \alpha \beta_2 p_{12})^2}{N,t}{2}+\frac{(\alpha-\alpha_N)\beta_1 ^2 }{4\alpha}\rQBav{q_{12}^2}{N,t}{2}=\\
 &=\frac1{2N} \rQBav{\mc Q_1}{N,t}1+\frac{\alpha_N}{2\alpha N}\rQBav{\mc R_{12}}{N,t}{2}+\frac{(\alpha-\alpha_N)\beta_1 ^2 }{4\alpha}\rQBav{q_{12}^2}{N,t}{2}.
\end{es}
Now, plugging Eqs. \eqref{eq:phi1}, \eqref{eq:phi0} and \eqref{eq:computation3} into Eq. \eqref{eq:GuerraPropagation}, we get the claim. The inequality \eqref{eq:HopfieldDecompositionIneq} then follows by noticing that the quantity $\frac12 \rQBav{\mc Q_1+\frac{\alpha_N}{\alpha}\mc R_{12}}{N,t}2$ is non-negative, and recalling that $\alpha_N = \max ( \lfloor \alpha N \rfloor, 1)/N$, the last contribution in the second line of Eq.~\eqref{eq:computation3} is non-negative as well for $N > \alpha^{-1}$.
\end{proof}

\begin{Rmk}\label{FinalBound}
The interpolation scheme introduced in Def.~\ref{InterMilan} yields a sharp bound on the Hopfield free energy in terms of a suitable combination of spin-glass free energies, since the associated remainder term is proven to be positive definite.  This improves upon the standard Guerra interpolation scheme \cite{Guerra2}, which, when applied to the Hopfield model under the replica-symmetric ansatz, leads to a remainder term whose sign is not controlled, in such a way that, neglecting such a term provides only an approximation of the free energy, rather than a rigorous bound.
\end{Rmk}

\red{
In the next Corollary, we generalize the specific choice $K(N)=\max(1,\lfloor \alpha N \rfloor)$, providing a lower bound for the intensive pressure valid for any extensive sequence $K'(N)$ yielding the same load $\alpha$ in the thermodynamic limit.
\begin{Cor}
    Let $K'(N)$ be any sequence of integers such that $\lim_{N\to\infty} K'(N)/N=\alpha>0$. Then, for all $\beta,\beta_1,\beta_2>0$ and $\psi\in \mR$, for $N>\alpha^{-1}$ we have
    \begin{ess}
        A_{N,K'(N)}(\beta) \ge A_N ^{SK} \big(\beta_1,\frac{\beta\psi}{\beta_1}\big)+\alpha_N  A_{\alpha_N N}^{sSK} (\beta_2,\beta) -\frac{\beta_1 ^2}{4}-\frac{\beta \psi^2}{2}-\frac{\beta}{2}\delta_N \mathbb E_{\bb\xi}[\lambda_{max}(\bb J)],
    \end{ess}
    where $\alpha_N = \max(1,\lfloor{\alpha N}\rfloor)/N$, $\delta_N = \frac{\overline K(N)- \underline K(N)}{\overline K(N)}$ where $\overline K(N)= \max(K'(N),\alpha_N N)$, $\underline K(N)= \min(K'(N),\alpha_N N)$ and $\lambda_{max}$ being the maximum eigenvalue of the Hebbian matrix.
\end{Cor}
\begin{proof}
    The proof works by joining the lower bound in Eq. \eqref{eq:HopfieldDecompositionIneq} and Rem. \ref{rmk:alpha_N}.
\end{proof}
}

\subsection{Streaming equations for the order parameters and their fluctuations}\label{4.2}
From a theoretical-physics perspective, the replica symmetry prescription is typically implemented by assuming that, in the thermodynamic limit and for any $t$, the probability distributions of the order parameters become Dirac deltas centered at their mean values, as a consequence, in Eq. \eqref{eq:GuerraPropagation}, the remainder $\partial_{t} \vphi_N (t)$ is trivially zero. Here we seek to bound this term properly, namely to prove that this is effectively vanishing under the milder SAP introduced in Def. \ref{def:weak_sap} and to check that this does hold true for whatever value of the interpolation parameter, particularly also at  $t=1$, where the standard statistical mechanical setting is recovered. 
\newline
To reach these goals we need the streaming-flow in $t$ of our interpolating structure, as provided in the next 

\begin{Prp}\label{prp:streamingO}
	Let $O:\hat \Sigma_N^{(s)}\to \mR$ be a function of $s$ replicas. Recalling that  $\mc Q_a = \beta N(m_a-\psi)^2$ and $\mc R_{ab}= \frac N2(\beta_1 q_{ab}-\sqrt{\alpha}\beta_2 p_{ab})^2$, we can state that if $\beta_1 \beta_2 = \sqrt{\alpha} \beta$, then
  	\begin{es}\label{eq:GeneralStreaming}
		\partial_ t\rQBav{O}{N,t}{s}&= \frac12\rQBav{O \big(\sum_{a=1}^s \mc Q_a-s\mc Q_{s+1}\big) }{N,t}{s+1}
  +\frac{\alpha_N}{\alpha}\rQBav{O\big(-\sum_{a< b}^s \mc R_{ab}+s\sum_{a=1}^s\mc R_{a,s+1}-\frac{ s (s+1)}{2} \mc R_{s+1,s+2}\big)
  }{N,t}{s+2}\\
  &+ \frac{\alpha-\alpha_N}{2\alpha}\beta_1 ^2 N\rQBav{O  \big(-\sum_{a<b}^s q_{ab}^2+s\sum_{a=1}^s q_{a,s+1}^2-\frac{s(s+1)}{2}q_{s+1,s+2}^2 \big)}{N,t}{s+2}.
	\end{es}
\end{Prp}

\begin{proof}
The proof is rather lengthy but fairly standard, hence we will report only the key steps. We start with
\begin{es}\label{eq:dtO}
    \partial_t \rQBav{O}{N,t}{s}&= \frac{\beta N}{2}\sum_{a=1}^s \rQBav{O(m_a-\psi)^2}{N,t}{s}-\frac s2\rQBav{O(m_{s+1}-\psi)^2}{N,t}{s+1} +\sum_{a=1}^s\rQBav{O \chi_t ^{(a)}}{N,t}{s}-s\rQBav{O \chi_t ^{(s+1)}}{N,t}{s+1},\nonumber
\end{es}
with
\begin{es}
    \chi^{(a)}_t&= \frac1{2\sqrt t}\sqrt{\frac{\beta}{N }}\sum_{i\mu}\xi^\mu_i \sigma^a_i z_\mu ^a-\frac1{2\sqrt{1-t}}\big(\frac{\beta_1}{\sqrt{2N}}\sum_{ij}J_{ij}\sigma^a_i \sigma^a _j+\frac{\beta_2}{\sqrt{2K}} \sum_{\mu\nu} \bar J_{\mu\nu}z_\mu^a z_\nu ^a\big)\\&	-\frac\beta2 \sum_{\mu } (z_\mu^a)^2+\frac{\beta_2^2}{4K}\sum_{\mu\nu}(z_\mu^a)^2 (z_\nu^a)^2.\nonumber
\end{es}
To evaluate the first three contributions we iteratively use Wick-Isserlis lemma. This way, and proceeding as in the proof of Prop. \ref{lem:mainprop}, we can write
\begin{eqnarray}
    \sum_{a=1}^s \rQBav{O\chi^{(a)}_t}{N,t}{s}&=&-\frac{\alpha_N}{\alpha}\sum_{a<b }^s\rQBav{ O \mc R_{ab}}{N,t}{s}+\frac{\alpha _N s}{2\alpha}\sum_{a=1}^s \rQBav{O \mc R_{a,s+1}}{N,t}{s+1}-\frac{\beta_1^2 N s}{4}\rQBav{O}{N,t}{s}\nonumber\\
    &-&\frac{(\alpha-\alpha_N) \beta_1 ^2 N}{2\alpha}\sum_{a< b}\rQBav{O q_{ab}^2}{N,t}{s}+\frac{(\alpha-\alpha_N)s }{4\alpha}\beta_1^2 N \sum_{a=1}^s \rQBav{O q_{a,s+1}^2}{N,t}{s+1}
    ,\\
    \rQBav{O \chi^{(s+1)}_t}{N,t}{s+1}&=&-\frac{\alpha_N}{2\alpha}\sum_{a=1 }^s\rQBav{ O \mc R_{a,s+1}}{N,t}{s+1}+\frac{\alpha _N (s+1)}{2\alpha} \rQBav{O \mc R_{s+1,s+2}}{N,t}{s+2}-\frac{\beta_1^2 N }{4}\rQBav{O}{N,t}{s}\nonumber\\
    &-&\frac{(\alpha-\alpha_N) \beta_1 ^2 N}{4\alpha}\sum_{a=1}^s\rQBav{O q_{a,s+1}^2}{N,t}{s+1}+\frac{(\alpha-\alpha_N)(s+1) }{4\alpha}\beta_1^2 N  \rQBav{O q_{s+1,s+2}^2}{N,t}{s+2}.\qquad
\end{eqnarray}
Plugging these results in the expression of $\partial_t \rQBav{O}{N,t}{s}$ and using the definition of $\mc Q_a$, we get the claim.
\end{proof}

\begin{Cor}\label{cor:stream_bounds}
The following cases hold:
\begin{itemize}
    \item[i)] Given a function $O_1$ of a single replica, the streaming equation reads as
    \begin{es}
        \label{eq:streaming_1replica}
        \partial _t \rQBav{O_1}{N,t}{1}&=\frac12 \rQBav{O_1 \mc Q_1}{N,t}{1}-\frac12 \rQBav{O_1 \mc Q_2}{N,t}{2}+\frac{\alpha_N}{\alpha}\rQBav{O_1\mc  R_{12}}{N,t}{2}-\frac{\alpha_N}{\alpha}\rQBav{O_1 \mc R_{23}}{N,t}{3}\\
        &+\frac{\alpha-\alpha_N}{2\alpha}\beta _1 ^2 N \rQBav{O_1(q_{12}^2-q_{23}^2)}{N,t}{3}.
    \end{es}
    \item[ii)] Given a function $O_{12}$ of two replicas, the streaming equation reads as
    \begin{es}
        \partial_t \rQBav{O_{12}}{N,t}{2}&= \frac12\rQBav{O_{12}(\mc Q_1+\mc Q_2)}{N,t}{2}-\rQBav{O_{12}\mc Q_3}{N,t}{3}\\
        &-\rQBav{O_{12}\mc R_{12}}{N,t}{2}+2(\rQBav{O_{12}\mc R_{13}}{N,t}{3}+\rQBav{O_{12}\mc R_{23}}{N,t}{3})-3\rQBav{O_{12}\mc R_{34}}{N,t}{4}
        \\&
        +\frac{\alpha-\alpha_N}{2\alpha}\beta _1 ^2 N [-\rQBav{O_{12} q_{12}^2}{N,t}{2}+2(\rQBav{O_{12} q_{13}^2}{N,t}{3}+
        \rQBav{O_{12} q_{23}^2}{N,t}{3})-3 \rQBav{O_{12} q_{34}^2}{N,t}{4}]
        .
    \end{es}
    In particular, if $O_{12}$ is symmetric under replica-indices exchange (i.e. $O_{ab}=O_{ba}$ for every pair of replica indices $a,b$), then
	\begin{es}
		\label{eq:streaming_2replicas}
		\partial_t \rQBav{O_{12}}{N,t}{2}&=\rQBav{O_{12}\mc Q_1}{N,t}{2}-\rQBav{O_{12}\mc Q_3}{N,t}{3}
  -\rQBav{O_{12} \mc R_{12}}{N,t}{2}+4\rQBav{O _{12}\mc R_{13}}{N,t}{3}-3\rQBav{O_{12}\mc R_{34}}{N,t}{4}\\
  &+\frac{\alpha-\alpha_N}{2\alpha}\beta _1 ^2 N [-\rQBav{O_{12} q_{12}^2}{N,t}{2}+4\rQBav{O_{12} q_{13}^2}{N,t}{3}-3 \rQBav{O_{12} q_{34}^2}{N,t}{4}].
	\end{es}
\end{itemize}
\end{Cor}
\begin{proof}
	The proof is a direct consequence of Prop. \ref{prp:streamingO}. As for Eq. \eqref{eq:streaming_2replicas}, we also use  the symmetry of $O_{12}$ and the invariance of quenched averages w.r.t. relabeling of replica indices to prove that $\rQBav{O_{12}\mc Q_{1}}{N,t}{2}=\rQBav{O_{12}\mc Q_{2}}{N,t}{2}$  and $\rQBav{O_{12}\mc R_{13}}{N,t}{3}=\rQBav{O_{12}\mc R_{23}}{N,t}{3}$, and similarly $\rQBav{O_{12} q^2_{13}}{N,t}{3}=\rQBav{O_{12} q^2_{23}}{N,t}{3}$. 
\end{proof}

\begin{Rmk}\label{rem:shift_O}
    Because of the peculiar structure of the streaming equations obtained in Prop.~\ref{prp:streamingO}, the shift $O\to O-c$, with $c$ being any function of $t$, leaves the r.h.s. of \eqref{eq:GeneralStreaming} (and consequently those in Eqs. \eqref{eq:streaming_1replica} and \eqref{eq:streaming_2replicas}) unchanged. In particular, if the limit $\bar O_t = \lim_{N\to\infty}\rQBav{O}{N,t}{s}$ exists and it is finite for any $t\in [0,1]$, we can safely replace $O$ with $O -\bar O_ t$, thus expressing the derivative of the quenched average of the observable $O$ in terms of correlation functions involving its fluctuations around its mean value.
\end{Rmk}

\subsection{Control of the remainder and of the fluctuations of the order parameters}\label{4.3}
We now merge the results from the previous two sub-sections to conclude the global proof.
\begin{Prp}\label{prp:RQnbounds}
    Let $\beta_1 \beta_2 = \sqrt{\alpha} \beta$. Recalling that $\mc Q_a = \beta N(m_a-\psi)^2$, $\mc R_{ab}= \frac N2(\beta_1 q_{ab}-\sqrt{\alpha}\beta_2 p_{ab})^2$, and $M_O^{(r)}$ is the upper-bound of the fluctuations of the arbitrary observable $O$ (as defined in Def.~\ref{def:weak_sap}), we can state that
    \begin{itemize}
        \item [i)]  If $m_1$ satisfies SAP at level $r\in \mN$, \\ then for all $0< n \le \left\lfloor \frac r2 \right\rfloor$, for all $N \in \mN$ and for almost every $t\in[0,1]$:
        \begin{equation}
            \label{eq:boundQn}
             \rQBav{\mc Q_1 ^n}{N,t}{1} \le\beta ^n [(M_m^{(r)})^{\frac 1r}+\sqrt N \vert \bar m_t -\psi\vert ]^{2n}.
        \end{equation}
        \item[ii)] If $(q_{12},p_{12})$ satisfies SAP at level $r\in \mN$, \\ then for all $0<n \le  \left\lfloor \frac {r+2}4\right\rfloor $, for all $N \in \mN$ and for almost every $t\in[0,1]$:
        \begin{equation}
            \label{eq:boundRn}
             \rQBav{\mc R_{12} ^n}{N,t}{2} \le\frac1{2^n} [\beta_1(M^{(r)}_q)^{\frac1r}+\beta_2 (M^{(r)}_p)^{\frac1r}+{\sqrt { N}} \vert \beta_1 \bar q_t -\sqrt{\alpha}\beta_2 \bar p_t\vert ]^{2n}.
        \end{equation}
    \end{itemize}
\end{Prp}

\begin{proof}  
First, keeping in mind Def. \ref{CeOpm} (particularly \eqref{eq:mu-signal}), the expression of $\rQBav{\mc Q_1 ^n}{N,t}{1}$ in terms of the fluctuations $\mu_1$ reads as
$$
 \rQBav{\mc Q_1 ^n}{N,t}{1}=\beta^n  \rQBav{[\mu_1+\sqrt N(\bar m_t -\psi)]^{2n}}{N,t}{1}
$$
Since $\mc Q_1$ is a non-negative random variable, taking the absolute value and expanding the r.h.s. by means of the binomial theorem and the triangle inequality we get:
$$
 \rQBav{\mc Q_1 ^n}{N,t}{1}\le \beta^n \sum_{k=0}^{2n} \binom{2n}{k}\vert  \rQBav{\mu_1^k}{N,t}{1}\vert \cdot \vert \sqrt N(\bar m_t -\psi)\vert ^{2n-k}.
$$
Now, $\vert  \rQBav{\mu^k}{N,t}{1}\vert\le   \rQBav{\vert\mu\vert ^k}{N,t}{1}\le [M^{(r)}_m]^{\frac kr}$, with the last inequality being valid for all $t\in[0,1]$ and all $N$ since SAP holds for $r\ge k_{max}=2n$. Thus
$$
 \rQBav{\mc Q_1 ^n}{N,t}{1}\le \beta^n \sum_{k=0}^{2n} \binom{2n}{k} [({M^{(r)}_m })^{\frac 1r}]^k \sqrt N\vert \bar m_t -\psi\vert ^{2n-k}=\beta^n [({M^{(r)}_m })^{\frac 1r}+\sqrt N \vert \bar m_t -\psi\vert ]^{2n},
$$
which is the statement at point $i)$. The proof for $\mc R_{12}$ works similarly, but it is lengthier:
\begin{es}
    \rQBav{\mc R_{12}^n}{N,t}{2}&=2^{-n}\vert\rQBav{(\beta_1 \theta_{12}-\beta_2\rho_{12}+\sqrt{N}(\beta_1 \bar q_t -\sqrt{\alpha} \beta_2 \bar p_t ) )^{2n}}{N,t}{2}\vert \le
   \\
   &\le 2^{-n}\sum_{k=0}^{2n}\binom{2n}{k}\vert\rQBav{(\beta_1 \theta_{12}-\beta_2\rho_{12})^k}{N,t}{2}\vert \cdot(\sqrt N \vert \beta_1 \bar q_t -\sqrt{\alpha} \beta_2 \bar p_t\vert )^{2n-k}.\nonumber
\end{es}
Now
\begin{es}
\vert\rQBav{(\beta_1 \theta_{12}-\beta_2\rho_{12})^k}{N,t}{2}\vert &=\left\vert \sum_{l=0}^k\binom{k}{l}(-1)^{k-l}\beta_1 ^l \beta_2 ^{k-l}\rQBav{\theta_{12}^l \rho_{12}^{k-l}}{N,t}{2}\right\vert\le \sum_{l=0}^k\binom{k}{l}\beta_1 ^l \beta_2 ^{k-l}\vert \rQBav{\theta_{12}^l \rho_{12}^{k-l}}{N,t}{2}\vert=\\
&=\beta_1 ^k \vert\rQBav{\theta_{12}^k}{N,t}{2}\vert+\beta_2 ^k \vert\rQBav{\rho_{12}^k}{N,t}{2}\vert+\sum_{l=1}^{k-1}\binom{k}{l}\beta_1 ^l \beta_2 ^{k-l}\vert \rQBav{\theta_{12}^l \rho_{12}^{k-l}}{N,t}{2}\vert.
    \nonumber
\end{es}
By means of the Cauchy-Schwarz inequality, we have $\vert \rQBav{\theta_{12}^l \rho_{12}^{k-l}}{N,t}{2}\vert\le \sqrt{\rQBav{\theta_{12}^{2l}}{N,t}{2}\rQBav{\rho_{12}^{2(k-l)}}{N,t}{2}}$. The highest possible moment involved in the previous equation is achieved when $l$ (or equivalently $k-l$) attains the maximum value. This yields by putting $l=k_{max}-1=2n-1$, so that the highest moment in $\theta_{12}$ is $\rQBav{\theta_{12}^{2(2n-1)}}{N,t}{2}$. If SAP holds at $r\ge 2(2n-1)$, then $\vert \rQBav{\theta_{12}^l \rho_{12}^{k-l}}{N,t}{2}\vert\le (M^{(r)}_q)^{\frac lr}(M^{(r)}_p)^{\frac {k-l}r} $ for all $N$ and almost every $t\in[0,1]$. Now, since $2(2n-1)\ge 2n$ for all $n\in \mN$, SAP holding at order $r\ge 2(2n-1)$ it is enough to bound $\vert\rQBav{\theta_{12}^k}{N,t}{2}\vert\le \rQBav{\vert\theta_{12}\vert^k}{N,t}{2}\le (M^{(r)}_q)^{\frac kr}$ and $\vert\rQBav{\rho_{12}^k}{N,t}{2}\vert\le \rQBav{\vert\rho_{12}\vert^k}{N,t}{2}\le (M^{(r)}_p)^{\frac kr}$ for almost every $t\in[0,1]$ and for all $N\in \mN$. Then, it follows that 
\begin{es}
  \vert\rQBav{(\beta_1 \theta_{12}-\beta_2\rho_{12})^k}{N,t}{2}\vert&\le \beta_1 ^k (M^{(r)}_q)^{\frac kr}+\beta_2 ^k (M^{(r)}_p)^{\frac kr}  +\sum_{l=1}^{k-1}\binom{k}{l}\beta_1 ^l \beta_2 ^{k-l}(M^{(r)}_q)^{\frac lr}(M^{(r)}_p)^{\frac {k-l}r}= \\&=(\beta_1(M^{(r)}_q)^{\frac1r}+\beta_2 (M^{(r)}_p)^{\frac1r})^k,
\end{es}
for all $N$ and almost every $t\in[0,1]$, hence
\begin{es}
     \rQBav{\mc R_{12}^n}{N,t}{2}&\le 
     2^{-n}\sum_{k=0}^{2n}\binom{2n}{k} (\beta_1(M^{(r)}_q)^{\frac1r}+\beta_2 (M^{(r)}_p)^{\frac1r})^k (\sqrt N \vert \beta_1 \bar q_t -\sqrt{\alpha} \beta_2 \bar p_t\vert )^{2n-k}=\\
     &=
     2^{-n}(\beta_1(M^{(r)}_q)^{\frac1r}+\beta_2 (M^{(r)}_p)^{\frac1r}+\sqrt N \vert \beta_1 \bar q_t -\sqrt{\alpha} \beta_2 \bar p_t\vert)^{2n},
     \nonumber
\end{es}
leading to the statement at point $ii)$.

\end{proof}
We also need the following
\begin{Prp}    \label{prp:G_function}
Define
\begin{equation}
\label{eq:G_function}
  \mathcal G_s(\bb q^2) =\sqrt N \big(-\sum_{a<b}^s q_{ab}^2+s\sum_{a=1}^s q_{a,s+1}^2-\frac{s(s+1)}{2}q_{s+1,s+2}^2 \big).  
\end{equation}
If $q_{12}$ satisfies SAP at order $r=4$, then for all $s\in \mN$ there exists a constant $\tilde C_s$ such that for all $N\in \mN$ and for almost all $t \in [0,1]$
    $$
    \rQBav{(\mc G_s(\bb q^2))^2}{N,t}{s+2}\le 4\tilde C^2_s.
    $$
\end{Prp}
\begin{proof}
We express the $\mc G$ function in terms of the fluctuations $\theta_{ab}$ by using $q_{ab}=N^{-1/2} \theta_{ab} +\bar q_t$. In this case, it is easy to show that
\begin{equation}
\mc G_s(\bb q^2)= 
{2\bar q_t}\hat {\mc G}_s(\bb \theta)+\frac1{\sqrt N} \hat{\mc G}_s (\bb \theta^2),
\end{equation}
where
$$ \hat{\mc G}_s(\bb q^2)=\frac1{ \sqrt N}\mc G_s(\bb q^2)= -\sum_{a<b}^s q_{ab}^2+s\sum_{a=1}^s q_{a,s+1}^2-\frac{s(s+1)}{2}q_{s+1,s+2}^2,$$
and $\hat {\mc G}_s(\bb \theta)$ and $\hat {\mc G}_s(\bb \theta^2)$ are obtained by replacing $q_{ab}^2$ with $\theta_{ab}$ and $\theta_{ab}^2$ respectively. This implies that
\begin{equation}
     \rQBav{(\mc G_s(\bb q^2))^2}{N,t}{s+2}= \rQBav{ 4 \bar q_t ^2 \hat {\mc G}_s(\bb \theta)^2+\frac{4 \bar q _t }{\sqrt N}\hat {\mc G}_s(\bb \theta)\hat {\mc G}_s(\bb \theta^2)+\frac 1N \hat {\mc G}_s(\bb \theta^2)^2}{N,t}{s+2}.
\end{equation}
Now, $\hat {\mc G}_s(\bb \theta)^2$, $\hat {\mc G}_s(\bb \theta)\hat {\mc G}_s(\bb \theta^2)$ and $\hat {\mc G}_s(\bb \theta^2)^2$ are polynomial in the fluctuations $\theta_{12}$ of order 2, 3 and 4 respectively:
\begin{eqnarray}
\hat {\mc G}_s(\bb \theta)^2&=& P_1 (\theta_{ab}^2, \theta_{ab}\theta_{cd}),\\
\hat {\mc G}_s(\bb \theta)\hat {\mc G}_s(\bb \theta^2)&=& P_2 (\theta_{ab}^3, \theta_{ab}^2\theta_{cd}),\\
    \hat {\mc G}_s(\bb \theta^2)^2&=& P_3 (\theta_{ab}^4, \theta_{ab}^2\theta_{cd}^2).
\end{eqnarray}
Crucially, $\hat {\mc G}_s(\bb \theta^2)^2$ does not depend on mixed products such as $\theta_{ab}^3 \theta_{cd}$, thus assuming SAP at order 4 for the order parameter $q_{12}$ is sufficient to bound all of these functions by combining again triangle and Cauchy-Schwarz inequalities, so that there exist constants $C_1^{(s)},$ $C_2^{(s)}$ and $C_3^{(s)}$ such that
\begin{eqnarray}
\rQBav{\hat {\mc G}_s(\bb \theta)^2}{N,t}{s+2}&\le & C_1 ^{(s)},\\
\vert\rQBav{\hat {\mc G}_s(\bb \theta)\hat {\mc G}_s(\bb \theta^2)}{N,t}{s+2}\vert&\le& C_2 ^{(s)},\\
\rQBav{\hat {\mc G}_s(\bb \theta^2)^2}{N,t}{s+2}&\le& C_3^{(s)},
\end{eqnarray}
for all $N$ and almost every $t$. It follows that
\begin{equation}
     \rQBav{(\mc G_s(\bb q^2))^2}{N,t}{s+2}\le  4 \bar q_t ^2C_1 ^{(s)}+\frac{4 \bar q _t }{\sqrt N}C_2 ^{(s)}+\frac { C_3 ^{(s)}}N,
\end{equation}
yielding the claim since $N>1$ and $\bar q_t\le 1,$ upon identifying $\tilde C^2_s=   C_1 ^{(s)}+C_2 ^{(s)}+\frac { C_3 ^{(s)}}4$.
\end{proof}

The following Lemma guarantees that, in the present formulation of Guerra's framework,  by suitably selecting the function $\psi$ and the noise $\beta_1$, the whole interpolation effectively becomes a morphing between the Hopfield neural network and a linear combination of hard and soft spin glasses: in particular, we will find out that  $\psi$ has to play the role of the Mattis magnetization. 

\begin{Lem}\label{lem:mqp}
Let $(\alpha,\beta)\in \mR^2_+$. There exists a choice of $\psi$ and $\beta_1$ with $\beta_1 \beta_2=\sqrt{\alpha}\beta$ such that, if $(m_1,q_{12},p_{12})$ satisfy the SAP at order four, then $\bar m_t =\bar m_{t=0}=:\bar m$, $\bar q_t = \bar q_{t=0}=:\bar q$ and $\bar p_t = \bar p_{t=0}=:\bar p$ for any $t\in [0,1]$. In such a case $(\psi,\beta_1)$ are identified as the solutions of the system\footnote{The fixed point equations \eqref{eq:fixedpoint_RS}, as well as SAP at the fourth order, match exactly the hypotheses appearing in the main Theorem of this part of the manuscript, that is Thm. \ref{thm:thermolim}.}\footnote{We stress that, although we force $\langle m \rangle_{N,t}, \langle q \rangle_{N,t}, \langle p \rangle_{N,t}$ to converge to the values $\bar m,\bar q, \bar p$ as $N \to \infty$, yet these do not need to be constant, e.g. $\bar m$ can still be a function of the storage and the noise.} 
\begin{equation}\label{eq:fixedpoint_RS}
     \begin{cases}
        \psi = \bar m (\psi,\beta_1),\\
        \beta_1 ^2 \bar q (\psi,\beta_1 )= {{\alpha\beta \bar p (\psi,\beta_1)}}
     \end{cases}.
\end{equation}
\end{Lem}
\begin{proof}
Let us start by fixing $\alpha,\beta,\beta_1$ studying the equation $\psi = \bar m_t (\psi,\beta_1)$. First, let us recall that $\vert\bar m_t\vert \le 1 $ always, so that $\bar m_t ([-1,1],\beta_1)\subseteq [-1,1] $ for all $\beta_1>0$. 
Since it is also $\psi$-continuous, by Brouwer's fixed point theorem there always exists at least one fixed point $\psi=\psi_t$. Notice that $\psi=0$ is a solution for all $t\in[0,1]$, then -- calling $t \mapsto \mc F_t$ the set-valued map associating with each $t$ the corresponding set of fixed points of $\psi=\bar m_t$ -- we trivially have $\cap_{t\in[0,1]} \mc F_t\neq \emptyset$. We then focus on common fixed points $\psi=\bar m_t(\psi,\beta_1)$ for all $t\in[0,1]$. For convenience, let us define $r_{12}=\beta_1 q_{12}-\sqrt{\alpha}\beta_2 p_{12},$ such that by SAP $\rQBav{r_{12}}{N,t}{2}\to \beta_1 \bar q_t -\sqrt{\alpha}\beta_2 \bar p_t:= \bar r_t$. According to Cor. \ref{cor:stream_bounds}, we have
\begin{ess}
    	\partial_t \rQBav{r_{12}}{N,t}{2}&=\rQBav{r_{12}(\mc Q_1-\mc Q_3)}{N,t}{3}
  +\frac{\alpha_N}{\alpha}\rQBav{r_{12}(- \mc R_{12}+4\mc R_{13}-3\mc R_{34})}{N,t}{4}+\frac{\alpha-\alpha_N}{2\alpha}\beta _1 ^2 \sqrt N \rQBav{r_{12}\mc G_2 (\bb q^2)}{N,t}{4}.
\end{ess}
Now, according to Rem. \ref{rmk:VIP}, we have
\begin{ess}
    \rQBav{r_{12}(\mc Q_1-\mc Q_3)}{N,t}{3}&=\beta\sqrt N (\bar m -\psi) \rQBav{r_{12} (\mu_1 -\mu_3)}{N,t}{2}+\frac{\beta}{2}\rQBav{r_{12}(\mu_1^2-\mu_3^2)}{N,t}{3}=\frac{\beta}{2}\rQBav{r_{12}(\mu_1^2-\mu_3^2)}{N,t}{3},
\end{ess}
since we fixed $\psi =\bar m_t$. Further, using the notation $\delta_{ab}= \beta_1 \theta_{ab}-\sqrt\alpha\beta_2 \rho_{ab}$, we can write
\begin{ess}
    \rQBav{r_{12}(- \mc R_{12}+4\mc R_{13}-3\mc R_{34})}{N,t}{4}&=\sqrt N \bar r_t \rQBav{r_{12}(-\delta_{12}+4\delta_{13}-3\delta_{34})}{N,t}{4}\\
    &+\frac12 \rQBav{r_{12}(-\delta_{12}^2+4\delta_{13}^2-3\delta_{34}^2)}{N,t}{4}.
\end{ess}
Using these results, we have
\begin{ess}
    	&\partial_t \rQBav{r_{12}}{N,t}{2}-\sqrt N \bar r_t \rQBav{r_{12}(-\delta_{12}+4\delta_{13}-3\delta_{34})}{N,t}{4}=\\
        =&\frac{\beta}{2}\rQBav{r_{12}(\mu_1^2-\mu_3^2)}{N,t}{3}
  +\frac{\alpha_N}{2\alpha}\rQBav{r_{12}(-\delta_{12}^2+4\delta_{13}^2-3\delta_{34}^2)}{N,t}{4}+\frac{\alpha-\alpha_N}{2\alpha}\beta _1 ^2 \sqrt N [\rQBav{r_{12}\mc G_2 (\bb q^2)}{N,t}{4}].
\end{ess}
Now, by SAP at order 4 again thanks to Rem. \ref{rmk:VIP} we have
\begin{eqnarray*}
    \lim_{N\to\infty}\rQBav{r_{12}(\mu_1 ^2-\mu_3^2)}{N,t}{3}&=&\lim_{N\to\infty}\rQBav{r_{12}}{N,t}2(\rQBav{\mu_1 ^2}{N,t}1-\rQBav{\mu_3^2}{N,t}{1})=0,\\
    \lim_{N\to\infty}\rQBav{r_{12}(-\delta_{12}^2+4\delta_{13}^2-3\delta_{34}^2)}{N,t}{4}&=& \lim_{N\to\infty}\rQBav{r_{12}}{N,t}2(-\rQBav{\delta_{12}^2}{N,t}{2}+4\rQBav{\delta_{13}^2}{N,t}2-3\rQBav{\delta_{34}^2}{N,t}2)=0,
\end{eqnarray*}
for almost every $t$. On the other hand, by Prop. \ref{prp:G_function}, we have
$$
\vert  \rQBav{r_{12}\mc G_2 (\bb q^2)}{N,t}{4}\vert \le \sqrt{\rQBav{r_{12}^2}{N,t}{2}}\sqrt {\rQBav{G_2 (\bb q^2)^2}{N,t}{4}}\le 2\tilde C_2 \sqrt{\rQBav{r_{12}^2}{N,t}2},
$$
with r.h.s. having finite limit, as $\rQBav{r_{12}^2}{N,t}{2}- (\rQBav{r_{12}}{N,t}{2})^2\to 0$ and $\lim_N \rQBav{r_{12}}{N,t}{2}= \beta_1 \bar q_t -\sqrt \alpha \beta_2 \bar p_t$. This means that 
$$
\lim_{N\to\infty}\frac{\alpha-\alpha_N}{2\alpha}\beta_1^2 \sqrt N\rQBav{r_{12}\mc G_2 (\bb q^2)}{N,t}{4}=0,
$$
since with $\alpha_N =\max(1,\lfloor{\alpha N}\rfloor)/N$ one has $\sqrt N (\alpha-\alpha_N)\to 0$. Using these results 
\begin{ess}
    \lim_{N\to\infty}\lVert \partial_t \rQBav{r_{12}}{N,t}{2}-\sqrt N \bar r_t \rQBav{r_{12}(-\delta_{12}+4\delta_{13}-3\delta_{34})}{N,t}{4}\lVert _\infty=0.
\end{ess}
Now, using the fact that $\sqrt N (r_{12}-\bar r_t)= \delta_{12}$, for some $C>0$ we have
$$
\sqrt N\vert \rQBav{r_{12}(-\delta_{12}+4\delta_{13}-3\delta_{34})}{N,t}{4}\vert =\vert \rQBav{\delta_{12}(-\delta_{12}+4\delta_{13}-3\delta_{34})}{N,t}{4}\vert \le C,
$$
again thanks to SAP and Cauchy-Schwarz inequality. In other words, $\lim_{N\to\infty}\lVert \partial_t \rQBav{r_{12}}{N,t}{2}- \bar r_t g_{N}(t)\lVert _\infty=0$ with $\sup_N \lVert g_N(t)\lVert_\infty \le C.$ 
As a result, we can extract a subsequence $g_{N_k}$ and a bounded function $g$ such that $g_{N_k}\rightharpoonup^* g$ in $L_\infty([0,1])$ by Banach-Alaoglu, and $\lim_{k\to\infty}\lVert \partial_t \rQBav{r_{12}}{N_k,t}{2}- \bar r_t g_{N_k}(t)\lVert _\infty=0$. Now, since $\rQBav{r_{12}}{N_k,t}{2}\to \bar r_t$
in $L_\infty$ norm (since $\rQBav{r_{12}}{N,t}{2}$ does) and $\bar
r \cdot 1_{[0,t]}\in L^1$, integrating the last limit between $0$ and $t$ gives $\bar r_t=\bar r_{t=0}+\int_0^t ds\,\bar r_s g(s)$, so that $\partial_t \bar r_t= \bar r_t g(t)$ for almost any $t$.
Solving the differential equation leads to $\bar r_t = \bar r_{t=0} \exp \int_0 ^t dt' g(t')$, or, more explicitly:
$$
\beta_1 \bar q_t -\sqrt{\alpha}\beta_2 \bar p_t=[\beta_1 \bar q_{t=0} -\sqrt{\alpha}\beta_2 \bar p_{t=0}]\cdot \exp \int _0 ^t dt' g(t').
$$
Then, if we choose $\beta_1$ as the zero of the equation $\beta_1 \bar q_{t=0} = \sqrt\alpha  \beta_2 \bar p_{t=0}$, it will be a zero for $\beta_1 \bar q_{t} = \sqrt \alpha \beta_2 \bar p_{t}$ for almost any $t$:
once this important result has been established, we can now focus on the quenched expectation of the magnetization. First, for all $t\in[0,1]$
$$
\vert\rQBav{m_1}{N,t}{1}-\rQBav{m_1}{N,0}{1}\vert\le t\cdot \esssup_{s\in[0,t]}\vert \partial_s \rQBav{m_1}{N,s}{1}\vert . 
$$
According again to Cor. \ref{cor:stream_bounds}, with SAP at order 4, we have 
\begin{es}
     \partial _s \rQBav{m_1}{N,s}{1} &=\frac12 \rQBav{m_1(\mc Q_1-\mc Q_2)}{N,s}{2} +\frac{\alpha_N }{\alpha} \rQBav{m_1(\mc R_{13} -\mc R_{23})}{N,s}{3}+\frac{\alpha-\alpha_N}{2\alpha}\beta_1^2 N \rQBav{m_1 \mc G_1 (\bb q^2)}{N,s}{3}.
\end{es}
Proceeding as before and using $\psi=\bar m_t$ and $\beta_1 \bar q_t =\sqrt \alpha \beta_2 \bar p_t$ for almost any $t$, we have
\begin{eqnarray*}
    \lim_{N\to\infty} \rQBav{m_1(\mc Q_1-\mc Q_2)}{N,s}{2}&=& \lim_{N\to\infty} \frac\beta2\rQBav{m_1(\mu_1^2-\mu_2^2)}{N,s}{2}=\lim_{N\to\infty}\frac\beta 2\rQBav{m_1}{N,s}{1}(\rQBav{\mu_1^2}{N,s}{2}-\rQBav{\mu_2^2}{N,s}{2})=0,\\
    \lim_{N\to\infty} \rQBav{m_1(\mc R_{13}-\mc R_{23})}{N,s}{3}&=&\lim_{N\to\infty}\frac12 \rQBav{m_1(\delta_{13}^2-\delta^2_{23})}{N,s}{3}=\lim_{N\to\infty}\frac12  \rQBav{m_1}{N,s}{1}(\rQBav{\delta_{13}^2}{N,s}{2}-\rQBav{\delta^2_{23}}{N,s}{2})=0,
\end{eqnarray*}
while
$$
\lim_{N\to\infty}\frac{\alpha-\alpha_N}{2\alpha}\beta_1^2 \sqrt N\rQBav{m_1\mc G_1 (\bb q^2)}{N,s}{3}=0,
$$
as $\mc G_1(\bb q^2)$ has bounded second moment and $\sqrt N (\alpha-\alpha_N)\to 0$. Then, $\lim_{N\to\infty}\esssup_{s\in[0,t]}\vert \partial_s \rQBav{m_1}{N,s}{1}\vert=0$, so that $\lim_{N\to\infty} \vert \rQBav{m_1}{N,t}{1}- \rQBav{m_1}{N,0}{1}\vert =0$ for all $t\in [0,1]$, that is $\bar m_t= \bar m_{t=0}$. With similar arguments, we can show that $\bar q_t = \bar q_{t=0}$ and $\bar p_t = \bar p_{t=0}$ for all $t\in [0,1]$.
\newline 
As a consequence, the choice of $\psi$ and $\beta_1$ are precisely identified by the system \eqref{eq:fixedpoint_RS}, once $\beta_1\beta_2 = \sqrt{\alpha}\beta $ is used. This concludes the proof. 
\end{proof}

\noindent
Summarizing, the previous Lemma states that,  
\begin{itemize}
    \item apart for an irrelevant shift in the temperature $\beta$, the external field $\psi$ appearing in the hard Sherrington-Kirkpatrick sector of the interpolation (see Theorem \ref{thm:thermolim} or Definition \ref{InterMilan})
    plays exactly the role of the Mattis magnetization in the thermodynamic limit.

    \item in order to represent the Hopfield model in terms of two spin-glasses one needs to set
    \begin{es}\label{eq:qp_const}
        \beta_1 \bar q =\sqrt \alpha \beta_2 \bar p,
    \end{es}
 as this correctly matches the related order parameters in the thermodynamic limit, see \cite{barra2012glassy}. 
\end{itemize}

\subsection{A new variational principle for the intensive quenched statistical pressure}\label{4.4}  

Now we are ready to prove the main theorem.

\begin{proof}
    [Proof of Thm. \ref{thm:thermolim}]
   Let us rewrite the equality \eqref{eq:HopfieldDecomposition} in Prop. \ref{lem:mainprop} as
    \begin{es}
    \label{eq:thm1_eq0}
         &A_{N,K}(\beta)-\big[ A_N ^{SK}\big(\beta_1,\frac{\beta\psi}{\beta_1}\big)+\alpha_N A_K ^{sSK}(\beta_2,\beta)-\frac{\beta_1 ^2}{4}-\frac{\beta \psi^2}{2}\big]=\\=&\int _0 ^1 ds\, \Big(\frac1{2N}\rQBav{\mc Q_1}{N,s}1+\frac{\alpha_N}{2\alpha N}\rQBav{ \mc R_{12}}{N,s}2+\frac{(\alpha-\alpha_N)\beta_1 ^2 }{4\alpha}\rQBav{q_{12}^2}{N,s}{2}\Big).
    \end{es}
In order to inspect the asymptotic behavior of $A_{N,K}(\beta)$ it is convenient to pose 
$$
F_N(\psi,\beta_1) = -  A_N^{SK}\big(\beta_1,\frac{\beta\psi}{\beta_1}\big)-\alpha _N A_K^{sSK}(\beta_2,\beta)+\frac{\beta_1 ^2}{4}+\frac{\beta \psi^2}{2}.
$$
Indeed, the asymptotic behavior of the sequence $F_N (\psi,\beta_1)$
is the same as the one in the square brackets in Eq. \eqref{eq:thm1_eq0}, however, the reversed sign allows us to work with global minima rather than global maxima, which is technically more convenient. In particular, we first aim to establish the existence of global minimizers of $F_N (\psi,\beta_1)$.
For all $N$, the function $F_N(\psi,\beta_1)$ is continuous on $\mR\times (0,\infty)$, as the log-partition functions are continuous (respectively in the parameters $(\beta_1, h= \beta\psi/\beta_1)$ for the hard spin glass sector and $\beta_2 = \sqrt{\alpha}\beta/ \beta_1$ for the soft one), while for the last two terms, namely $\frac{\beta_1 ^2}{4}$ and $\frac{\beta \psi^2}{2}$, continuity is trivial.  

In order to prove the existence of global minima of $F_N(\psi,\beta_1)$, we show that it grows super-linearly at the boundary of the relevant domain of $(\psi,\beta_1)$. In particular, we claim that
$$
{F_N(\psi,\log \beta_1)}\to +\infty, \quad \text{as}\quad \lVert (\psi, \log\beta_1)\lVert \to\infty.
$$
To show this to be true, we  must control the two terms $A_N ^{SK}\big(\beta_1,\frac{\beta\psi}{\beta_1}\big)$ and $A_K^{sSK}(\beta_2,\beta)$ and check how they behave asymptotically. Let us start with the first:  because of super-additivity, we have $A_N^{SK} (\beta_1,\frac{\beta \psi}{\beta_1})\ge  A_1 ^{SK}(\beta_1,\frac{\beta \psi}{\beta_1})\ge 0$, since $A_1 ^{SK}(\beta_1,\frac{\beta \psi}{\beta_1})=\log [2\cosh (\beta \psi)]\ge 0$. Further, it is straightforward to see that 
    $A_N^{SK}(\beta_1,\frac{\beta\psi}{\beta_1})\le \beta_1 e_0 +\log 2\cosh(\beta \psi)$, with $e_0 \approx 0.763$ being the ground-state energy of the SK model in the thermodynamic limit, namely the limit of $e_0^{(N)}=- \frac 1N \mE{\bb J} \min_{\bb\sigma\in\Sigma_N} H_{N,\bb J ,h=0} ^{SK}(\bb\sigma)$ (which is super-additive sequence \cite{guerra2002thermodynamic}, so that $\lim_N e^{(N)}_0 = \sup_N  e^{(N)}_0 $).
   Putting the two pieces together, one has
    \begin{equation}\label{eq:coerc_SK}
0\le A_N^{SK}\big(\beta_1,\frac{\beta \psi}{\beta_1}\big)\le  {\beta_1 e_0}+ \log [2\cosh (\beta \psi)].
    \end{equation}
Let us now focus on the second term, namely $A_K^{sSK}(\beta_2,\beta)$. Let us recall from Rem. \ref{rmk:limit_beta} that $A_K ^{sSK}(\beta_2,\beta) \le A(\beta_2,\beta,p)$ for any positive function $p$ of $\beta_2,\beta$. Setting $\beta_2 =\sqrt{\alpha}\beta/\beta_1$, in order to make explicit the dependence on $\beta_1$, the expression of $p$ minimizing the functional $A(\beta_2,\beta,p)$ is 
\begin{es}\label{eq:p_solution}
  \beta<1:    p=
\begin{cases}
  \frac{\beta_1 (\sqrt \alpha \beta -\beta_1 (1-\beta))}{\alpha \beta^2},   \quad \beta_1<\frac{\sqrt \alpha \beta}{1-\beta} \\
    0,\quad\quad\quad\quad\quad \quad\quad\,\, \beta_1 >\frac{\sqrt \alpha \beta}{1-\beta}\end{cases},\quad \beta>1: p=\frac{\beta_1 (\sqrt \alpha \beta -\beta_1 (1-\beta))}{\alpha \beta^2}  {\forall\beta_1>0}.
\end{es}
By plugging this expression in the functional $A(\beta_2,\beta,p)$, regardless of the values of $\beta$ and $\alpha$ we get
$$
\lim_{\beta_1\to 0^+}A(\beta_2,\beta, p) =-\infty,
$$
in agreement with Rem. \ref{rmk:limit_beta}. Overall we have $\lim_{\beta_1\to\infty} A(\beta_2,\beta,p)=-\frac12\log (1-\beta)$ for $\beta <1$, and
\begin{es}\label{eq:sSK_asymptotic}
   A(\beta_2,\beta,p) = \frac{(1-\beta)^2 \beta_1^2}{4\alpha\beta^2}+\mc O (\beta_1), 
\end{es}
for large $\beta_1$ if $\beta>1$. Using Eq. \eqref{eq:coerc_SK} and Rem. \ref{rmk:limit_beta}, for $N>\alpha^{-1}$ we get $F_N (\psi,\beta_1) \ge g(\psi,\beta_1),$
with 
$$g(\psi,\beta_1)=-\beta_1 e_0 -\log[2\cosh(\beta \psi)]-\alpha A(\beta_2,\beta,p)+\frac{\beta_1^2}{4}+\frac{\beta \psi^2}{2},$$ with $p$ satisfying Eq. \eqref{eq:p_solution}. We now use the decomposition $g(\psi,\beta_1) = g_1 (\psi)+g_2(\beta_1,\beta)$, with $g_1(\psi)=\tfrac12 \beta\psi^2 -\log [2\cosh(\beta \psi)]$ and $g_2(\beta_1,\beta)= \tfrac14 \beta_1 ^2 -\beta_1e_0 -\alpha A(\tfrac{\sqrt \alpha \beta}{\beta_1},\beta,p)$ and analyze the asymptotic behavior of each contribution separately.
\par\medskip

{\it Asymptotic behavior of $g_1$.} Since $\log[2\cosh(\beta \psi)]= \beta \vert \psi\vert +\mc O(1)$ for $\vert \psi\vert \to\infty$, by Young's inequality there exist constants $c_1,C_1>0$ such that
$$
g_1(\psi) \ge c_1 \psi^2 -C_1\quad \forall \psi \in \mR.
$$
In particular, $g_1(\psi)\to+\infty$ for $\vert \psi\vert\to\infty$.
\par\medskip

{\it Asymptotic behavior of $g_2$.} From the limiting behavior of the soft SK sector (provided by Rem. \ref{rmk:limit_beta} and Eqs. (\ref{eq:p_solution},\ref{eq:sSK_asymptotic})), it follows that there exist constants $c_2,C_2,c_2',C_2'>0$ and $\gamma>0$ such that
\begin{equation}
    g_2(\beta_1) \ge \begin{cases}
        c_2 \beta_1 ^2 -C_2,\quad \text{for large } \beta_1,\\
        c_2 '\beta_1 ^{-\gamma} -C_2 ',\quad \text{for small }\beta_1.
    \end{cases}
    \nonumber
\end{equation}
Then, $g_2 (\beta_1)\to+\infty$ for $\beta_1\to 0^+$ and $\beta_1\to+\infty$.
\par\medskip
Let $K_R=\{(\psi,\beta_1): \sqrt{\psi^2+\beta_1^2}\le R\}$ and $K_R^c$ the complementary set. If $(\psi_n,\beta_{1,n})$ with $\lVert (\psi_n,\beta_{1,n})\lVert \to\infty$, then elements of the sequence will eventually lie outside $K_R$, namely $\vert \psi_n \vert >R$ or $\vert \log \beta_{1,n} \vert >R$ for sufficiently large $n$. In this case, $g_1(\psi_n)\to+\infty$ or $g_2(\beta_{1,n},\beta)\to+\infty$, thus implying $g(\psi_n,\beta_{1,n})\to+\infty$. Because of the arbitrariness of $R$ and calling $\zeta_1 = \log \beta_1$, we must conclude that
\begin{es}\label{eq:coercivity_new}
    \lim_{\lVert(\psi,\zeta_1)\lVert\to \infty}g(\psi,e^{\zeta_1})=+\infty,
\end{es}
that is, $g$ is coercive. Since $F_N (\psi,\beta_1)\ge g(\psi,\beta_1)$ eventually, it immediately follows that $F_N$ is itself coercive for sufficiently large $N$.
Now, by Guerra-Toninelli scheme applied on each SK sector, the function $F_N$ has point-wise limit
$$
F(\psi,\beta_1) := \lim_{N \to \infty}F_N(\psi,\beta_1) = -A^{SK}\big(\beta_1,\frac{\beta\psi}{\beta_1}) -\alpha A^{sSK}(\beta_2,\beta) +\frac{\beta_1^2}4+\frac{\beta\psi^2}{2}.
$$
Since the point-wise limit of each SK sector is the $N$-supremum of the sequence, we immediately have $F_N\ge F$ for $N>\alpha^{-1}$. Further, for all $N\in\mathbb N$, besides the continuity of the models $A^{SK}_N$ and $A_K^{sSK}(\beta_2,\beta)$, their derivatives w.r.t. the external field $\psi$ and the thermal noise $\beta_1$ are written in terms of quenched expectations of the magnetization and the overlaps, which are continuous and bounded (uniformly in $N$) on any compact subset $C$ of the domain, i.e. there exists a constant $M_C$ such that for all $N$
$$
\lVert \nabla A_N ^{SK}\lVert \le M_C,\quad \text{ for all } (\psi, \beta_1)\in C,
$$
and similarly for the soft-SK sector. Because of the above results, both the SK sectors are uniformly bounded in $C$, and by boundedness of the above derivatives, the sequences are further equi-continuous. Merging Arzelà-Ascoli theorem with point-wise convergence for the SK models, we immediately have uniform convergence of $A_N^{SK}$ and $A_K^{sSK}$ on any compact set $C$. Since it is a linear combination of (equi)continuous sequences converging uniformly in any
compact set, the $F_N$ itself will be an equicontinuous family of functions (and by this, the limit $F$ is itself continuous) converging uniformly on any compact sets. Since $F$ is continuous in any compact $C$, it is continuous over the whole domain.
\newline
Now, since $F_N\ge g$ for all $N>\alpha^{-1}$, then $F\ge g$, so that the family $F_N$ and $F$ are (resp. eventually) coercive. Since they are also continuous, they attain their global minima in some compact set 
$$
K_\lambda = \{(\psi,\beta_1)\in \mR\times (0,\infty): g(\psi,\beta_1)\le \lambda\},
$$
for some $\lambda\in \mR.$ Let us define for convenience $X= \mathcal{R}\times (0,\infty)$, and call $b_N = \min_X F_N $, $b=\min_X F$ and $B=\operatorname{argmin}_X F$. Since $F_N\ge F $ eventually, we have $b_N = \min_X F_N \ge \min _X F =b$ for $N>\alpha^{-1}$; this implies $\liminf_{N\to\infty} b_N \ge b$. On the other hand, given $x^*\in B$, we have $b_N =\min_X F_N \le F_N (x^*)$. Due to the pointwise convergence $F_N(x^*)\to F(x^*)=b$, we have $\limsup_{N\to\infty} b_N \le b$. Joining the two results, we thus have
\begin{equation} \label{eq:convergenceFN}
\exists \lim_{N\to\infty}\min_X F_N = \lim_{N\to\infty}b_N = b= \min_X F,
\end{equation}
which provides the convergence of global minima of $F_N$ to the global minimum of $F$.
\par
We now go back to Eq. \eqref{eq:thm1_eq0} and take the minimum over $\psi,\beta_1$, from which we have
    \begin{es}\label{eq:equality_toberef}
    &    \min_{\beta_1 >0,\psi\in \mR}\big[A_{N,K}(\beta)+(-A_N ^{SK}\big(\beta_1,\frac{\beta\psi}{\beta_1}\big)-{\alpha_N} A_K ^{sSK}(\beta_2,\beta)+\frac{\beta_1 ^2}{4}+\frac{\beta \psi^2}{2})\big]=\\=    &\min_{\beta_1 >0,\psi\in \mR}\int _0 ^1 ds\, \Big(\frac1{2N}\rQBav{\mc Q_1}{N,s}1+\frac{\alpha_N}{2\alpha N}\rQBav{ \mc R_{12}}{N,s}2+\frac{(\alpha-\alpha_N)\beta_1 ^2 }{4\alpha}\rQBav{q_{12}^2}{N,s}{2}\Big).
    \end{es}
    while for $N>\alpha^{-1}$ 
    \begin{es}
     \min_{\beta_1 >0,\psi\in \mR}\big[  A_{N,K}(\beta)&+(-A_N ^{SK}\big(\beta_1,\frac{\beta\psi}{\beta_1}\big)-{\alpha_N} A_K ^{sSK}(\beta_2,\beta)+\frac{\beta_1 ^2}{4}+\frac{\beta \psi^2}{2})\big]\ge 0,
    \end{es}
    since in the r.h.s. of Eq. \eqref{eq:thm1_eq0} we only have non-negative quantities.
    According to Lem. \ref{lem:mqp}, we choose $\psi= \psi^*$, $\beta_1 =\beta_1 ^*$ being the solutions of the constraints $\psi=\bar m(\psi,\beta_1)$ and $\beta_1^2 \bar q(\psi,\beta_1)=\alpha\beta \bar p(\psi,\beta_1)$. If  now $(m_1,q_{12},p_{12})$ satisfy SAP at order $4$, by Prop. \ref{prp:RQnbounds} we can bound the r.h.s. of Eq. \eqref{eq:equality_toberef} as 
    \begin{es}
    \label{eq:chain}
        &     \min_{\beta_1 >0,\psi\in \mR} \int _0 ^1 ds\Big(\frac1{2N}\rQBav{\mc Q_1}{N,s}1+\frac{\alpha_N}{2\alpha N}\rQBav{ \mc R_{12}}{N,s}2+\frac{(\alpha-\alpha_N)\beta_1 ^2 }{4\alpha}\rQBav{q_{12}^2}{N,s}{2}\Big)\le\\
       &\le   \int _0 ^1 ds\Big(\frac1{2N}\rQBav{\mc Q_1}{N,s}1+\frac{\alpha_N}{2\alpha N}\rQBav{ \mc R_{12}}{N,s}2+\frac{(\alpha-\alpha_N)\beta_1 ^2 }{4\alpha}\rQBav{q_{12}^2}{N,s}{2}\Big)\Big\vert_{(\psi,\beta_1)=(\psi^*,\beta_1^*)}
        \le \\
          & \le \frac1{2N} [\beta \sqrt{M_m ^{(4)}}+\frac{\alpha_N}{2\alpha}  (\beta_1^*(M_q^{(4)})^{\frac14}+\frac{\sqrt{\alpha} \beta}{\beta_1^*} (M_p^{(4)})^{\frac14})^2]+\frac{(\alpha-\alpha_N)\beta_1 ^2 }{4\alpha}\rQBav{q_{12}^2}{N,s}{2}\Big\vert_{(\psi^*,\beta_1^*)}.
    \nonumber\end{es}
   With these premises, the following chains of inequalities holds:
    \begin{es}\label{eq:incremental}
   0& \le  A_{N,K}(\beta)+\min_{\beta_1 >0,\psi\in \mR}\big(-  A_N ^{SK}\big(\beta_1,\frac{\beta\psi}{\beta_1}\big)-\frac{\alpha_N}\alpha(\alpha  A_K ^{sSK}(\beta_2,\beta)-\frac{ \beta_1 ^2}{4})+\frac{\beta \psi^2}{2}\big)\le
      \\&\le\frac1{2N}
   \big(\beta \sqrt{M_m ^{(4)}}+\frac{\alpha_N}{2\alpha}[\beta_1^*(M_q^{(4)})^{\frac14}+\frac{\sqrt{\alpha} \beta}{\beta_1^*} (M_p^{(4)})^{\frac14}]^2\big)+\frac{(\alpha-\alpha_N)\beta_1 ^2 }{4\alpha}\rQBav{q_{12}^2}{N,s}{2}\Big\vert_{(\psi^*,\beta_1^*)}.\nonumber
    \end{es}
The r.h.s. in the previous line vanishes in the thermodynamic limit, in such a way that
\begin{es}\label{eq:limit_of_difference}
  \lim_{N\to\infty}\Big[A_{N,K}(\beta)+\min_{\beta_1 >0,\psi\in \mR}\big(-  A_N ^{SK}\big(\beta_1,\frac{\beta\psi}{\beta_1}\big)-\alpha_N  A_K ^{sSK}(\beta_2,\beta)+\frac{ \beta_1 ^2}{4}+\frac{\beta \psi^2}{2}\big)\Big]=0.  
\end{es}
Thus, because of \eqref{eq:convergenceFN}, the limit of $A_{N,K}(\beta)$ exists, and reads
\begin{es}
    A(\alpha,\beta)=\lim_{N\to\infty}A_{N,K}(\beta)&=\max_{\beta_1>0,\psi\in \mR}\big[ A ^{SK}\big(\beta_1,\frac{\beta\psi}{\beta_1}\big)+\alpha A ^{sSK}(\beta_2,\beta)-\frac{\beta \psi^2}{2}-\frac{\beta_1 ^2}{4}\big],
\end{es}
from which the statement is proved. Because of Rem. \ref{rmk:alpha_N}, this is true for all $K(N)$ such that $K(N)/N\to \alpha$ in the thermodynamic limit.
\end{proof}

We can finally show that under SAP hypothesis, the  expression for the intensive quenched statistical pressure of the Hopfield model in the thermodynamic limit exists and it is exactly the one provided by AGS theory \cite{AGS1,AGS2}, that is, once proved that the variational formulation of the problem as captured by Theorem \ref{thm:thermolim} is well posed, we intend to find its solution, thereby proving Corollary \ref{thm:AGS}, i.e. the AGS formula.

\begin{proof}
    [Proof of Cor. \ref{thm:AGS}]
The expression of the quenched pressures for the Sherrington-Kirkpatrick model and its soft version under the SAP hypothesis at $t=0$ are given by the replica-symmetric formulas:
 \begin{eqnarray}
 	A^{SK}\big(\beta_1,\frac{\beta \psi}{\beta_1}\big)&=&\min_{ q\in[0,1]}\big(\int _\mR d\mu(\tau)\log2\cosh(\beta \psi+\beta_1 \tau\sqrt{ q})+\frac{\beta_1^2}{4}(1- q)^2\big),	\label{eq:pressure_SK_RS}\\
 	A^{sSK}(\beta_2,\beta)&=&\min_{ p\in \mR_+}\Big( \frac{ \beta_2 ^2}2 \frac{ p}{1-\beta+\beta^2_2  p}+\frac{\beta_2 ^2}{4} p ^2-\frac12 \log (1-\beta+\beta_2^2  p)\Big)\label{eq:pressure_sSK_RS},
 \end{eqnarray}
 with $\beta_2 = \sqrt \alpha \beta/\beta_1$. Then, using Thm. \ref{thm:thermolim}, we can recast the pressure of the Hopfield model as
 \begin{equation}\label{eq:maxmin}
     A(\alpha,\beta)= \max_{\beta_1>0,\psi \in \mR}\min_{q\in [0,1],p\in\mR_+}\big( A ^{SK}\big(\beta_1,\frac{\beta\psi}{\beta_1}\big)+\alpha  A ^{sSK}\big(\frac{\sqrt \alpha \beta}{\beta_1},\beta\big)-\frac{\beta_1 ^2}{4}-\frac{\beta \psi^2}{2}\big).
 \end{equation}
 The extremization w.r.t. $q$ leads to the self-consistency equation
 \begin{equation}
     q = \int _\mR d\mu(\tau) \tanh^2 (\beta\psi+\beta_1 \tau\sqrt q),\label{eq:q_cons_RS}
     \end{equation}
     for the SK model while the stationary condition w.r.t. $p$ is satisfied by the solution
     \begin{equation}
     \label{eq:p_cons_RS}
     p=\begin{cases}
         0\quad\quad\quad\quad\quad \beta_2 <1-\beta\\
         \frac{\beta_2-(1-\beta)}{\beta_2^2}\quad\quad \beta _2 >1-\beta
     \end{cases},
\end{equation}
if $\beta<1$, and $p=\frac{\beta_2-(1-\beta)}{\beta_2^2}$ if $\beta>1$ in the soft sector. In the latter case $p=0$ is no longer an admissible stationary point, since the domain constraint $1-\beta+\beta_2^2 p>0$ underlying $A^{sSK}$ forces $p>(\beta-1)/\beta_2^2>0$, so that $p=\frac{\beta_2-(1-\beta)}{\beta_2^2}$ is the only solution left. In terms of the SK thermal noise, the constraints in Eq. \eqref{eq:p_cons_RS} for $\beta_2$ are $\beta_1>\sqrt \alpha \beta/(1-\beta)$ and $\beta_1<\sqrt \alpha \beta/(1-\beta)$ respectively. By requiring the stationarity condition w.r.t. $\psi$ we get the further constraint
\begin{eqnarray}
    \psi = \int _\mR d\mu (\tau) \tanh (\beta \psi+\beta_1 \tau\sqrt q).\label{eq:psi_cons_RS}
\end{eqnarray}
Finally, the derivative of the argument of the max-minimization problem in Eq. \eqref{eq:maxmin} can be written in full generality as
$$
\partial_{\beta_1 }\big( A ^{SK}\big(\beta_1,\frac{\beta\psi}{\beta_1}\big)+\alpha  A ^{sSK}(\beta_2,\beta)-\frac{\beta_1 ^2}{4}-\frac{\beta \psi^2}{2}\big)= -\frac{\beta_1 q^2}{2}+\frac{\alpha^2 \beta^2}{2\beta_1 ^3}p ^2.
$$
Now, if $\beta <(1+\sqrt \alpha)^{-1} $ (which is $<1$), it is possible to find  $\beta_1\in (\frac{\sqrt \alpha\beta}{1-\beta},1)$ such that $p=0$ by Eq. \eqref{eq:p_cons_RS} and $q=0$ are the only solution to Eq. \eqref{eq:q_cons_RS}, so that $\partial_{\beta_1 }\big( A ^{SK}\big(\beta_1,\frac{\beta\psi}{\beta_1}\big)+\alpha  A ^{sSK}(\beta_2,\beta)-\frac{\beta_1 ^2}{4}-\frac{\beta \psi^2}{2}\big)=0$ trivially: this is the solution in the ergodic region, which also implies $\psi =0$ by Eq. \eqref{eq:psi_cons_RS}. Conversely, if $\beta >(1+\sqrt \alpha)^{-1}$, we can look for $\beta_1 <\frac{\sqrt \alpha\beta}{1-\beta}$ such that $p>0$. In that case, we can write
$$
-\frac{\beta_1 q^2}{2}+\frac{\alpha^2 \beta^2}{2\beta_1 ^3}p ^2= -\frac1{2\beta_1 }(\beta_1 ^2 q^2 -\frac{\alpha^2 \beta^2}{\beta_1 ^2}p ^2)= -\frac1{2\beta_1 }(\beta_1 q-\frac{\alpha\beta}{\beta_1}p)(\beta_1 q+\frac{\alpha\beta}{\beta_1}p).
$$
Then, the only possibility to have a vanishing $\beta_1$-derivative of the objective function in Eq. \eqref{eq:maxmin} is to set 
$$\beta_1^2 =\frac{\alpha\beta p}q,$$
which is the same fixed point problem studied in Lem. \ref{lem:mqp}. Focusing our attention on the $q,p>0$ case, we get
$$
A^{sSK}(\beta_2,\beta)= \frac{\beta q}{2[1-\beta(1-q)]}+\frac{\beta_1^2q^2}{4\alpha}-\frac12 \log[1-\beta(1-q)].
$$
Putting all the pieces together in Eq. \eqref{eq:maxmin}, we finally get
\begin{es}
    A(\alpha,\beta)&= -\frac{\beta \psi^2}{2}+\int _\mR d\mu(\tau)\log \left( 2\cosh(\beta \psi+\tau\sqrt{\alpha \beta p}) \right) -\frac{\alpha\beta}{2}p (1-q)\\&+\frac\alpha 2 \frac{ \beta q}{1-\beta(1-q)}-\frac\alpha2 \log[1-\beta(1-q)],
\end{es}
with $\psi,p,q$ being the solutions of the equations
\begin{eqnarray}
    \psi &=& \int _\mR d\mu (\tau) \tanh (\beta \psi+ \tau\sqrt {\alpha\beta p}),\\
    q &=& \int _\mR d\mu (\tau) \tanh ^2(\beta \psi+ \tau\sqrt {\alpha\beta p}),\\
    p &=&\frac{\beta q}{[1-\beta(1-q)]^2}.
\end{eqnarray}
These expressions precisely reproduce  the Amit-Gutfreund-Sompolinsky replica-symmetric picture for the Hopfield model as claimed.
\end{proof}

\section{Discussion and conclusions}\label{SezioneCinque}

\red{In this work we contributed to address a long-standing problem in the mathematical theory of mean-field neural networks, namely the existence of the thermodynamic limit of the quenched free energy of the Hopfield model in the high-storage regime. The main result of the present paper is to show that, under suitable concentration assumptions on the relevant order parameters, the existence of the infinite-volume limit can be established rigorously. 
\newline
More broadly, our results provide two main contributions to the mathematical physics of neural networks, one conceptual and one technical: from a conceptual viewpoint, the key ingredient is a novel Guerra-like interpolation scheme that avoids dealing directly with the Hopfield Hamiltonian. Indeed, by leveraging the equivalence between Hopfield networks and restricted Boltzmann machines and the representation of the latter in terms of a suitable combination of hard and soft Sherrington–Kirkpatrick spin glasses, we construct an interpolating framework connecting the original neural network to spin-glass models whose thermodynamic limits are already known. While Guerra’s original interpolations provide powerful bounds for standard spin glasses, their direct application to neural networks has so far led to approximations rather than rigorous bounds: here, instead, the remainder entering the sum rule has a definite sign, thus neglecting it results in a replica-symmetric expression providing a genuine bound on the quenched free energy rather than an uncontrolled approximation.
\newline
From a more technical viewpoint, rather than assuming self-averaging by postulating that the distributions of the order parameters collapse to Dirac deltas at their means, we explicitly control how this concentration has to occur. In particular, by introducing suitable weak self-averaging properties, we quantify the fluctuations around the limiting means and 
find a degree of concentration sufficient for the remainder of the interpolation to vanish in the thermodynamic limit: this strengthens, within the Guerra interpolation framework, the spirit of the approach pioneered by Pastur, Shcherbina and Tirozzi and leads to the rigorous existence of the thermodynamic limit in the replica-symmetric regime. 
\newline
As a consequence, the corresponding variational principle emerges naturally and yields precisely the AGS expression for the free energy. The assumptions employed in this work are deliberately tailored to the replica-symmetric regime. Nevertheless, the interpolation itself is not intrinsically restricted to this setting. Since the representation of restricted Boltzmann machines in terms of hard and soft spin glasses is already available beyond replica symmetry, at least at the level of the first step of replica symmetry breaking, it is natural to expect that the present strategy can be extended to more complex overlap structures: such developments could provide a route toward a rigorous Parisi-type theory for the Hopfield model.
\newline
Furthermore, the methodology introduced here is not restricted to the classical Hopfield network: the underlying decomposition through bipartite spin-glass representations suggests possible applications to a wider class of associative-memory architectures, including dense and higher-order Hopfield networks, random-feature models, Boltzmann machines and neural networks endowed with consolidation or dreaming mechanisms. We therefore view the present work as a first step toward a general interpolation framework for establishing thermodynamic limits in a broad family of mean-field neural networks. 
\newline
Establishing the existence of the free energy without any concentration hypothesis remains an important open problem, and we hope that the combination of interpolation techniques, weak self-averaging and spin-glass representations developed here will provide useful ingredients toward its eventual resolution.}

\section*{Acknowledgments}
A.B. and E.A. acknowledge financial support from Sapienza University of Rome, project number RM12519999AB8CA9 (``Neural Networks and Learning Machines: asymptotic behaviors on structured datasets").
\newline
E.A. and A.F. acknowledge financial support from PNRR MUR project PE0000013-FAIR and from Sapienza University of Rome (RM120172B8066CB0, RM12218169691087, RM123188F3CD7763, RM124190CB1269EB).  
\newline
D.P. acknowledges financial support from PNRR MUR project number 20223L2NWK (``Elliptic and parabolic problems, heat kernel estimates and spectral theory").
\newline
The Authors acknowledge useful discussions with L. Albanese and A. Alessandrelli.

\appendix
    \section{\red{Universality of quenched noise}}\label{app:universality}

\red{In this Appendix, we provide a rigorous argument supporting the validity of the universality hypothesis of the quenched noise, which allows us to describe the Hopfield model with binary patterns in terms of an auxiliary one consisting of a Mattis ferromagnetic contribution -- accounting for the signal -- plus a sea of Gaussian patterns contributing to the noise. This will provide a further incremental step of universality results \cite{Carmona,Genovese}. We stress that, for the purposes of the present manuscript, this property is established {\it only} under the assumption of condensation of a single pattern (which is later formalized), that is, a single pattern is candidate to be retrieved at the equilibrium. This of course does not cover the full phenomenology of the Hopfield model, but still it provides a clear framework to inspect universality of quenched noise. Our strategy again relies on a Guerra interpolation bridging the two models, the core of the argument lying in the control of the remainder of the resulting sum rule. To encode the single-pattern condensation hypothesis, we adopt a {\it leave-one-out} prescription: the interpolation involves non-condensed patterns only, while the one which is candidate to be retrieved at the equilibrium is left untouched. Let us fix the notation: for consistency with the main text, we denote by $\bb\eta$ the Rademacher retrieved pattern, and call $\bb \eta^1,\dots,\bb\eta^K\in \{+1,-1\}^N$ the remaining binary ones (we are thus considering the Hopfield model with $K+1$ Boolean patterns $\bb\eta,\bb\eta^1,\dots,\bb\eta^K$). On the other hand, Gaussian patterns are denoted by $\xi^\mu_i \underset{i.i.d}\sim\mc N(0,1)$. To better stress the difference between the two models, we use the superscript $H$ for the Hopfield networks with binary patterns. Then, the Hamilton functions of the two models read as
\begin{es}
   H^{(H)}_{N,K,\bb\eta}(\bb\sigma)=-\frac N2 m_{\bb\sigma}^2-\frac N{2}\sum_{\mu=1}^K \Big(\frac 1N {\sum_{i=1}^N}\eta^\mu_i  \sigma_i\Big)^2,
\end{es}
for the Hopfield model, while the mixed Mattis-Gaussian one is given in Eq. \eqref{eq:HamiltonHopfield}, which we report here for the sake of completeness:
\begin{es}
   H_{N,K,\bb\eta,\bb \xi}(\bb\sigma)=-\frac N2 m_{\bb\sigma}^2-\frac N{2}\sum_{\mu=1}^K \Big(\frac 1N {\sum_{i=1}^N}\xi^\mu_i  \sigma_i\Big)^2.
\end{es}
The structure of Guerra's interpolation is fixed by the following
\begin{Def}
    Consider the process on $\Sigma_N$:
    \begin{es}
            Y_\mu ^{(t)}(\bb\sigma) = \sqrt{t}\frac 1N {\sum_{i=1}^N}\eta^\mu_i  \sigma_i+\sqrt{1-t}\frac 1N {\sum_{i=1}^N}\xi^\mu_i  \sigma_i.
    \end{es}
    The interpolating partition function is
    \begin{es}
        Z_{N,K,\bb\eta,\bb\xi,t} = \sum_{\bb\sigma \in \Sigma_N}\exp\Big(\frac{\beta N}2 m_{\bb\sigma}^2+\frac {\beta N}{2}\sum_{\mu=1}^K ( Y_\mu ^{(t)}(\bb\sigma))^2\Big),
    \end{es}
    with associated interpolating pressure $A_{N,K}(t) = \frac1N \mathbb E_{\bb\eta,\bb\xi}\log  Z_{N,K,\bb\eta,\bb\xi,t}$ and Boltzmann average $\omega_{N,t}(\cdot)$.
\end{Def}}
\red{
The interpolating model is itself of Hopfield-type with generalized stored patterns $\hat \eta^\mu_i (t)= \sqrt t \eta^\mu_i + \sqrt{1-t} \xi^\mu_i$ and associated overlaps $Y_\mu^{(t)} (\bb\sigma) = \frac1N \sum_{i=1}^N \hat\eta^\mu_i(t) \sigma_i$, with the Hamilton function still retaining the Hebbian structure with these interpolating vectors: $H_{N,K,\bb\eta,\bb\xi,t}(\bb\sigma) =-\frac N2 m_{\bb\sigma}^2 -\frac1{2N}\sum_{\mu=1}^K\sum_{i,j=1}^N\hat \eta^\mu_i (t)\hat \eta^\mu_j (t)\sigma_i \sigma_j$. As a consequence, the partition function still admits an RBM representation \cite{BarraEquivalenceRBMeAHN} with a single set of hidden variables (this would not be the case if one interpolated between the two quadratic contributions separately, since the Boltzmann Machine representation would require two sets of hidden units).}
\par\medskip
\red{
We now proceed to formalize the assumption of single-pattern condensation. The crucial observation is that, in the regime of interest, the free energy landscape is dominated (at large $N$) by pure states which are strongly correlated with the retrieved pattern only and essentially uncorrelated with the remaining ones. Accordingly, the fields $Y_\mu^{(t)}$ asymptotically behave as Gaussian random variables with variance of order $1/N$, so that their quenched moments of order $n$ scale as $N^{-n/2}$ (or, more precisely, can be bounded by $D^{(n)}/N^{n/2}$, for some constant $D^{(n)}>0$). As we now show, non-condensation of patterns making up the noisy bulk in the high-storage regime can be encoded by the scaling of the $4$-th moment alone. To this aim, we focus on the Hopfield model ($t=1$), where the Gaussian patterns $\bb\xi$ play no role. Then, for all (small but fixed) $0<c<1$, by Markov's inequality, for all $\mu =1,\dots, K$:
$$
\mathbb P (\vert m_\mu \vert \ge c )\equiv \mathbb E_{\bb\eta}\omega_{N,t=1}(1_{\vert m_\mu\vert \ge c}) \le \frac{\mathbb E_{\bb\eta}\omega_{N,t=1}(m_\mu^4)}{c ^4}\le \frac{D^{(4)}}{c^4 N^2}.
$$
By union bound, we have
$$
\mathbb P (\exists \mu=1,\dots,K : \vert m_\mu \vert \ge c) \le \sum_{\mu=1}^K\mathbb P (\vert m_\mu \vert \ge c ) \le \frac{\alpha _N D^{(4)}}{c^4 N}\to 0.
$$
Conversely, using constraints on the second moment, the above argument would provide an upper bound $\frac{\alpha _N D^{(2)}}{c^2}=\mc O(1)$ for $\mathbb P (\exists \mu=1,\dots,K : \vert m_\mu \vert \ge c)$, which is insufficient to ensure the non-condensation of the patterns $\bb\eta^\mu$. Hence, the fourth moment is the lowest one whose scaling beats the extensive number of non-retrieved patterns. To better inspect the scaling behavior of the 4-th moment for generic $t\in [0,1]$, we give the following
\begin{Prp}\label{prp:universality_ergo}
    For $\beta <1$, there exists a constant $C(\beta) <\infty$ such that
    $$
    \mathbb E_{\bb\eta,\bb\xi}\omega_{N,t}((Y_\mu^{(t)}(\bb\sigma))^4)\le \frac{C(\beta)}{N^2}, \quad\quad \forall \mu=1,\dots,K,
    $$
    for all $t \in [0,1]$ and for all $N\in \mathbb N$.
\end{Prp}
\begin{proof}
    Let us define $\tilde Y _\mu^{(t)}(\bb\sigma)=\sqrt N Y_\mu^{(t)}(\bb\sigma) $. From the definition of the Boltzmann average, we can write
    $$
    \omega_{N,t}((\tilde Y_\mu^{(t)}(\bb\sigma))^4)= \frac{\omega_{N,t}^{(\mu)} ((\tilde Y_\mu^{(t)}(\bb\sigma))^4\exp(\tfrac\beta2( \tilde Y_\mu^{(t)}(\bb\sigma))^2))}{\omega_{N,t}^{(\mu)} (\exp(\tfrac\beta2( \tilde Y_\mu^{(t)}(\bb\sigma))^2))}\le \omega_{N,t}^{(\mu)} ((\tilde Y_\mu^{(t)}(\bb\sigma))^4\exp(\tfrac\beta2(\tilde Y_\mu^{(t)}(\bb\sigma))^2)),
    $$
    with $\omega_{N,t}^{(\mu)}$ being the Boltzmann average after removal of the $\mu$-th field $\tilde Y_\mu^{(t)}(\bb\sigma)$. The last line follows since the argument of the exponential is non-negative, thus the denominator is always greater than 1. Averaging w.r.t. the quenched disorder:
    \begin{es}\label{eq:passo1}
       \mathbb E_{\bb\eta,\bb\xi}\omega_{N,t}((\tilde Y_\mu^{(t)}(\bb\sigma))^4)&\le \mathbb E_{\bb \eta_{\neg \mu},\bb\xi_{\neg \mu}}\omega_{N,t}^{(\mu)} (\mathbb E_{\bb\eta^\mu,\bb\xi^\mu} [(\tilde Y_\mu^{(t)}(\bb\sigma))^4\exp(\tfrac\beta2 \tilde Y_\mu^{(t)}(\bb\sigma)^2)])\le \\
       &\le \sup_{\bb\sigma\in \Sigma_N} \mathbb E_{\bb\eta^\mu,\bb\xi^\mu} ((\tilde Y_\mu^{(t)}(\bb\sigma))^4\exp(\tfrac\beta2 \tilde Y_\mu^{(t)}(\bb\sigma)^2)).
    \end{es}
    Here, $\mathbb E_{\bb \eta_{\neg \mu},\bb\xi_{\neg \mu}}$ is the average w.r.t. the patterns other than the $\mu$-th one. Notice that the only dependence on the patterns $\bb\eta^\mu$ and $\bb\xi^\mu$ is captured by the inner function, as the Boltzmann average $\omega^{(\mu)}_{N,t}$ does not depend on them. This allows the interchangeability of $\omega_{N,t}^{\mu}$ and $\mathbb E_{\bb\eta^\mu,\bb\xi^\mu}$. The last line follows by upper bounding the Boltzmann average with the supremum over $\Sigma_N$, which removes the dependence of the argument on the patterns with label $\nu \neq \mu$. Now, the random variable $ \hat \eta^\mu_i (t)=\sqrt t \eta^\mu_i+\sqrt{1-t}\xi^\mu_i$ is sub-Gaussian, since for all $\lambda\in \mR$
    $$
    \mathbb E_{\eta^\mu_i,\xi^\mu_i} \exp(\lambda \hat \eta^\mu_i (t))=\mathbb E_{\eta^\mu_i}\exp(\lambda\sqrt t \eta^\mu_i ) \mathbb E_{\xi^\mu_i}\exp(\lambda\sqrt{1- t} \xi^\mu_i )=\cosh (\lambda \sqrt t)\exp(\tfrac{1-t}{2}\lambda^2)\le \exp(\tfrac{\lambda^2}{2}),
    $$
    and $\cosh (x)\le \exp(x^2/2)$. As a result and due to the independence of the entries, for fixed $\bb\sigma \in \Sigma_N$, $\tilde Y_\mu^{(t)}$ is sub-Gaussian itself:
    $$
    \mathbb E _{\bb\eta^\mu,\bb \xi^\mu}\exp(\lambda \tilde Y_\mu^{(t)}(\bb\sigma))= \prod_{i=1}^N\mathbb E _{\eta^\mu_i, \xi^\mu_i} \exp(\tfrac{\lambda \sigma_i \hat \eta^\mu_i (t)}{\sqrt N})\le \exp(\frac{\lambda^2 }{2N}\sum_i\sigma_i ^2)= \exp(\tfrac{\lambda^2 }{2}).
    $$
    With this result, we can control the tail behavior of $\tilde Y_\mu^{(t)}$. For all $u>0$, by the Chernoff bound we have
    $$
    \mathbb P (\tilde Y_\mu^{(t)}(\bb\sigma) \ge u)\le \inf_{\lambda>0} e^{-\lambda u}\mathbb E_{\bb\eta^\mu,\bb\xi^\mu}\exp(\lambda \tilde Y_\mu^{(t)}(\bb\sigma))\le \inf_{\lambda>0} \exp(-\lambda u+\tfrac{\lambda^2}{2})= \exp(-\tfrac {u^2}2).
    $$
    Similarly:
    $$
    \mathbb P (\tilde Y_\mu^{(t)}(\bb\sigma) \le -u)\le \inf_{\lambda<0} e^{\lambda u}\mathbb E_{\bb\eta^\mu,\bb\xi^\mu}\exp(\lambda \tilde Y_\mu^{(t)}(\bb\sigma))\le \inf_{\lambda<0} \exp(\lambda u+\tfrac{\lambda^2}{2})= \exp(-\tfrac {u^2}2).
    $$
    By union bound, we have $\mathbb P (\vert\tilde Y_\mu^{(t)}(\bb\sigma) \vert\ge u)\le 2 \exp(-\tfrac {u^2}2)$. Calling $h(x) = x^4 \exp(\tfrac{\beta x^2} 2 )$, we have
    \begin{ess}
       \mathbb E_{\bb\eta^\mu,\bb\xi^\mu} ((\tilde Y_\mu^{(t)}(\bb\sigma))^4\exp(\tfrac\beta2 \tilde Y_\mu^{(t)}(\bb\sigma)^2))=\mathbb E_{\bb\eta^\mu,\bb \xi^\mu}h( \tilde Y_\mu^{(t)}(\bb\sigma)).
    \end{ess}
    The argument of the expectation depends on the $\mu$-th patterns only through the law of the field $ \tilde Y_\mu^{(t)}(\bb\sigma)$. Since $h$ is even, $h(\vert x\vert)$ is increasing and $h(0)=0$, we can use the formula $\mathbb E h(X) = \int _0 ^\infty du\, h'(u) \mathbb P(\vert X\vert \ge u)$, yielding
\begin{ess}
    \mathbb E_{\bb\eta^\mu,\bb\xi^\mu} ((\tilde Y_\mu^{(t)}(\bb\sigma))^4\exp(\tfrac\beta2 \tilde Y_\mu^{(t)}(\bb\sigma)^2))&=\int _0 ^\infty du (4 u^3+\beta u^5)\exp(\tfrac{\beta u^2}{2}) \mathbb P( \vert\tilde Y_\mu^{(t)}(\bb\sigma)\vert\ge u)\le
    \\&\le 2\int _0 ^\infty du (4 u^3+\beta u^5)\exp(-\tfrac{(1-\beta) u^2}{2}).
\end{ess}
The integral converges if and only if $\beta <1$. Upon integration, we have
$$
\mathbb E_{\bb\eta^\mu,\bb\xi^\mu} ((\tilde Y_\mu^{(t)}(\bb\sigma))^4\exp(\tfrac\beta2 \tilde Y_\mu^{(t)}(\bb\sigma)^2))\le \frac{16}{(1-\beta)^2}+\frac{16\beta}{(1-\beta)^3}:= C(\beta).
$$
Merging these results, we finally have
$$\sup_{\bb\sigma\in\Sigma_N}\mathbb E_{\bb\eta^\mu,\bb\xi^\mu}h(\tilde Y_\mu^{(t)}(\bb\sigma))\le\frac{16}{(1-\beta)^2}+\frac{16\beta}{(1-\beta)^3}:= C(\beta).$$
Using $\tilde Y _\mu^{(t)}(\bb\sigma)=\sqrt N Y_\mu^{(t)}(\bb\sigma) $, we get $\mathbb E_{\bb\eta,\bb\xi}\omega_{N,t}(( Y_\mu^{(t)}(\bb\sigma))^4)\le C(\beta)/N^2$, proving the claim.
    \end{proof}
The above proposition allows us to define a proper way to formally characterize the assumption of single-pattern condensation. The argument developed there works for $\beta<1$, a region including the whole paramagnetic phase ($\beta<(1+\sqrt \alpha)^{-1}$) as well as a portion of the spin glass one. The restriction primarily stems from the rough step of bounding the denominator in Eq. \eqref{eq:passo1} from below by $1$. The line $\beta=1$ is then a threshold of the method rather than a meaningful physical transition, and it is reasonable to expect that the same conclusion holds throughout the spin-glass phase, possibly by means of sharper techniques (see e.g. \cite{TalaNN1}). The same expectation applies, more importantly, to the retrieval region: there, the ordered phase is encoded in the macroscopic behavior of the overlap with the pattern $\bb\eta$, while the fields $Y_\mu^{(t)}$ associated to non-retrieved patterns are expected to retain the same asymptotic behavior. This is precisely the content of the following
\begin{Def}[Non-condensation]\label{def:SPC}
Let $(\alpha,\beta)\in \mR_+^2$ and $K(N)$ such that $\lim_{N\to\infty} K(N)/N=\alpha>0$. For this choice of parameters, we say that the fields $Y_\mu^{(t)}$ are non-condensed if there exists a constant $C(\alpha,\beta)>0$ such that
\begin{es}
    \mathbb E_{\bb \eta,\bb \xi}\omega_{N,t}((Y_\mu^{(t)}(\bb\sigma))^4)\le \frac{C(\alpha,\beta)}{N^2},\quad\quad\quad \forall \mu=1,\dots,K,
\end{es}
uniformly in $t\in[0,1]$ and for $N$ large enough.
\end{Def}
\begin{Rmk}
Def.~\ref{def:SPC} formalizes the noise-related notion of the single-pattern condensation, namely the requirement that no pattern in the bulk is retrieved. That the retrieval takes place on $\bb\eta$ plays no role in the universality argument given below, and is not formalized here. We stress also that the distinction between the retrieved pattern $\bb\eta$ and non-retrieved ones $\bb\eta^\mu$ is purely conventional, and only serves to separate the signal from the noisy bulk: any of the $K+1$ Boolean patterns can play this role upon relabeling. Configurations aligned with several patterns simultaneously -- such as spurious states -- are excluded by Def.~\ref{def:SPC} and fall outside the scope of the present argument.
\end{Rmk}
\begin{Lem}[Approximate Stein's lemma for bounded random variables, adapted from \cite{Genovese}]\label{Genovese}
    Let $X$ be a real bounded random variable, $\vert X\vert \le M$, such that, if $Z\sim \mc N(0,1)$, $\mathbb E X^k = \mathbb E Z^k$ for $k=1,\dots,m$ with $m\ge 2$, and let $f\in C^m([-M,M])$ with $\lVert f^{(m)}\lVert_{\infty,M}= \sup_{\vert x \vert \le M}\vert f^{(m)}(x)\vert$. Then
    $$
    \vert \mathbb E X f(X) - \mathbb E[X^2] \mathbb E f'(X) \vert \le \Big(\frac{m+1}{m!}\Big)\mathbb E [\vert X\vert ^{m+1} ]\lVert f^{(m)}\lVert _{\infty,M}.
    $$
\end{Lem}
\begin{proof}
    The proof is identical to that in \cite{Carmona,Genovese}. The only difference lies in the range of the supremum at the r.h.s. This follows from the fact that Taylor--Lagrange remainders of $f$ and $f'$
around $0$ are evaluated at intermediate points $\zeta\in[-M,M]$, and all
the remaining terms only involve $f^{(k)}(0)$ and moments of $X$ up to order
$m+1$, which are finite by boundedness.
\end{proof}
}
\red{
\begin{Lem}[Universality]
Let $(\alpha,\beta) \in \mR_+^2$, and $K=K(N)$ such that, for $N\to\infty$, $\alpha_N = K(N)/N\to \alpha >0$. If the fields $Y^{(t)}_\mu$ are non-condensed at $(\alpha,\beta)$, then
$$
\lim_{N\to\infty}\vert A_{N,K}^{(H)}(\beta)- A_{N,K}(\beta)\vert =0,
$$
with $A_{N,K}^{(H)}(\beta)$ being the quenched pressure of the Hopfield model with binary patterns, and $A_{N,K}(\beta)$ introduced in Def. \ref{def:pressureH}.
\end{Lem}
\begin{proof}
  By construction, at $t=1$ the interpolating quenched pressure reproduces the Hopfield model with $K+1$ Rademacher patterns, while at $t=0$ it reduces to the model with a single binary pattern and a bulk of $K$ Gaussian vectors. By the fundamental theorem of calculus
  $$
  A^{(H)}_{N,K}(\beta)= A_{N,K}(\beta)+\int _0 ^1 ds \, \partial_s A_{N,K}(s),
  $$
  where $A_{N,K}(\beta)$ is the one introduced in Def. \ref{def:pressureH}. This directly implies that
  \begin{es}\label{bound_universality}
        \vert A^{(H)}_{N,K}(\beta)- A_{N,K}(\beta)\vert \le \int _0 ^1 ds \,\vert \partial_s A_{N,K}(s)\vert.
  \end{es}
    The explicit expression of the derivative is
    $$
    \partial_s A_{N,K}(s) =\frac{\beta }{2N}\sum_{i=1}^N \sum_{\mu=1}^K\Big[\frac{1}{\sqrt s}\mathbb E_{\bb \eta, \bb \xi}{\eta^\mu_i} \omega_{N,s}(Y_\mu^{(s)}(\bb\sigma)\sigma_i)-\frac{1}{\sqrt {1-s}}\mathbb E_{\bb \eta,\bb \xi}{\xi^\mu_i} \omega_{N,s}(Y_\mu^{(s)}(\bb\sigma)\sigma_i)\Big].
    $$
   For the first contribution, we use the tower rule to express $$\mathbb E_{\bb \eta,\bb \xi} \eta^\mu_i  \omega_{N,s} (Y^{(s)}_\mu (\bb\sigma) \sigma_i ) = \mathbb E_{\bb \eta_{\neg \{i,\mu\}}, \bb \xi} \mathbb E_{\eta^\mu_i}[\eta^\mu_i f(\eta^\mu_i)\vert \bb \eta_{\neg \{i,\mu\}}, \bb \xi ],$$
  where $\bb \eta_{\neg \{i,\mu\}}$ is obtained by the Rademacher patterns dropping the $\eta^\mu_i$ entry, and 
  $$  f(x ) = \omega_{N,s}^{(x)}(Y_\mu^{(s)}(\bb\sigma,x)\sigma_i),$$
  with $\bb \eta_{\neg \{i,\mu\}}$ and $\bb\xi$ fixed, and $\eta^\mu_i$ being everywhere replaced by a generic variable $x$ (in $Y_\mu^{(s)}$ and, consequently, in the Boltzmann average $\omega_{N,t}$). This should be thought of as a function of $x$ alone. For the second contribution we can directly use Stein's lemma. Then we get
\begin{ess}
    \partial_s A_{N,K}(s)& =\frac{\beta }{2N}\sum_{i=1}^N \sum_{\mu=1}^K\Big[\frac{1}{\sqrt s}\mathbb E_{\bb \eta_{\neg \{i,\mu\}}, \bb \xi} \mathbb E_{\eta^\mu_i}[f'(\eta^\mu_i)\vert \bb \eta_{\neg \{i,\mu\}}, \bb \xi ]-\frac{1}{\sqrt {1-s}}\mathbb E_{\bb \eta,\bb \xi}\partial_{\xi^\mu_i} \omega_{N,s}(Y_\mu^{(s)}(\bb\sigma)\sigma_i)\Big]\\
    &+\frac{\beta }{2N}\sum_{i=1}^N \sum_{\mu=1}^K\Big[\frac{1}{\sqrt s}\mathbb E_{\bb \eta_{\neg \{i,\mu\}}, \bb \xi} \big(\mathbb E_{\eta^\mu_i}[\eta^\mu_i f(\eta^\mu_i)\vert \bb \eta_{\neg \{i,\mu\}}, \bb \xi ]-\mathbb E_{\eta^\mu_i}[f'(\eta^\mu_i)\vert \bb \eta_{\neg \{i,\mu\}}, \bb \xi ]\big)\Big].
\end{ess}
The second contribution in the first line is
  \begin{ess}
  \mathbb E_{\bb\eta,\bb\xi}\partial_{\xi^\mu_i }\omega_{N,s}(Y_\mu^{(s)}(\bb\sigma)\sigma_i)=\frac{\sqrt{1-s}}{N}+\beta \sqrt{1-s}\mathbb E_{\bb\eta,\bb\xi}[\omega_{N,s}((Y_\mu ^{(s)})^2)-\omega_{N,s}(Y_\mu ^{(s)}\sigma_i)^2].
  \end{ess}
  With the same arguments, the first contribution in the same line is the same after replacing $\sqrt {1-s}$ with $\sqrt s$. Plugging these results in the expression of the derivative, the first line trivially cancels out. Taking the modulus, we have
  \begin{es}
       \vert\partial_s A_{N,K}(s)\vert& =\frac{\beta }{2N}\Big\vert\sum_{i=1}^N \sum_{\mu=1}^K\Big[\frac{1}{\sqrt s}\mathbb E_{\bb \eta_{\neg \{i,\mu\}}, \bb \xi} \big(\mathbb E_{\eta^\mu_i}[\eta^\mu_i f(\eta^\mu_i)\vert \bb \eta_{\neg \{i,\mu\}}, \bb \xi ]-\mathbb E_{\eta^\mu_i}[f'(\eta^\mu_i)\vert \bb \eta_{\neg \{i,\mu\}}, \bb \xi ]\big)\Big]\Big\vert\le\\
       &\le \frac{\beta }{2N\sqrt s}\sum_{i=1}^N \sum_{\mu=1}^K\mathbb E_{\bb \eta_{\neg \{i,\mu\}}, \bb \xi} \vert\mathbb E_{\eta^\mu_i}[\eta^\mu_i f(\eta^\mu_i)\vert \bb \eta_{\neg \{i,\mu\}}, \bb \xi ]-\mathbb E_{\eta^\mu_i}[f'(\eta^\mu_i)\vert \bb \eta_{\neg \{i,\mu\}}, \bb \xi ]\vert.
  \end{es}
  By Lem. \ref{Genovese} (dropping the conditioned variables to lighten the notation), we have
  $$
  \vert \mathbb E_{\eta^\mu_i}[\eta^\mu_i f(\eta^\mu_i)]-\mathbb E_{\eta^\mu_i}[ f'(\eta^\mu_i)]\vert \le \frac{2}{3}\lVert f''' \lVert_{\infty,1},
  $$
  since $\mathbb E_{\eta^\mu_i} (\eta^\mu_i)^k = \mathbb E_{Z\sim \mc N(0,1)} Z^k$ for $k=1,2,3$ (thus $m=3$ in the Lemma), and $\mathbb E ( \eta^\mu_i ) ^2=\mathbb E \vert \eta^\mu_i \vert ^4=1$. Then, $\vert\partial_s A_{N,K}(s)\vert\le \frac{\beta}{3N\sqrt s}\sum_{i,\mu}^{N,K}\mathbb E_{\bb \eta_{\neg \{i,\mu\}}, \bb \xi}\lVert f''' \lVert_{\infty,1}$. By straightforward but cumbersome computations, one finds
  \begin{ess}
    f'''(x)&=\beta^3 s^{3/2} [\omega^{(x)}_{N,s}((Y_\mu^{(s)}(\bb\sigma,x)) ^4)-4 \omega^{(x)}_{N,s}( (Y_\mu^{(s)}(\bb\sigma,x))^3\sigma_i) \omega^{(x)}_{N,s}( Y_\mu^{(s)}(\bb\sigma,x)\sigma_i)\\
    &-3\omega^{(x)}_{N,s}( (Y_\mu^{(s)}(\bb\sigma,x))^2)^2-6\omega^{(x)}_{N,s}( Y_\mu^{(s)}(\bb\sigma,x)\sigma_i)^4+12 \omega^{(x)}_{N,s}(Y_\mu^{(s)}(\bb\sigma,x) \sigma_i )^2 \omega^{(x)}_{N,s}( (Y_\mu^{(s)}(\bb\sigma,x))^2)] .
  \end{ess}
  Taking the modulus, we have the bound
\begin{ess}
    \vert f'''(x)\vert& \le \beta^3 s^{3/2} [\omega^{(x)}_{N,s}(\vert Y_\mu^{(s)}(\bb\sigma,x)\vert  ^4)+4 \omega^{(x)}_{N,s}( \vert Y_\mu^{(s)}(\bb\sigma,x)\vert ^3) \omega^{(x)}_{N,s}(\vert Y_\mu^{(s)}(\bb\sigma,x)\vert)\\
    &+3\omega^{(x)}_{N,s}( \vert Y_\mu^{(s)}(\bb\sigma,x)\vert^2)^2+6\omega^{(x)}_{N,s}( \vert Y_\mu^{(s)}(\bb\sigma,x)\vert )^4+12 \omega^{(x)}_{N,s}(\vert Y_\mu^{(s)}(\bb\sigma,x) \vert )^2 \omega^{(x)}_{N,s}( \vert Y_\mu^{(s)}(\bb\sigma,x)\vert ^2)].
\end{ess}
Now, by H\"older's inequality, for $k=1,2,3$ we have $\omega^{(x)}_{N,s}(\vert Y_\mu^{(s)}\vert ^k)\le \omega^{(x)}_{N,s}(\vert Y_\mu^{(s)}\vert ^4)^{\frac k4}$, so that
\begin{es}\label{eq:3rd_der}
    \vert f'''(x)\vert& \le 26 \beta^3 s^{3/2} \omega^{(x)}_{N,s}(( Y_\mu^{(s)}(\bb\sigma,x))  ^4).
\end{es}
Let us now better exploit the structure of the r.h.s. starting with the identity
\begin{es}
    \label{eq:why}
    Y_\mu^{(s)}(\bb\sigma,x) = \frac{\sqrt{s}}N (x-\eta^{\mu}_i)\sigma_i + Y_\mu^{(s)}(\bb\sigma).
\end{es}
The Boltzmann average is therefore
\begin{ess}
    \omega^{(x)}_{N,s}(( Y_\mu^{(s)}(\bb\sigma &,x))  ^4) =\frac{\sum_{\bb\sigma \in \Sigma_N}( Y_\mu^{(s)}(\bb\sigma,x))  ^4 \exp\Big(\frac{\beta N}2 m_{\bb\sigma}^2 +\frac{\beta N}{2}(Y_\mu^{(s)}(\bb\sigma,x))^2+\frac{{\beta N}}{2}\sum_{\nu \neq \mu}(Y_\nu^{(s)}(\bb\sigma))^2 \Big) 
    }{\sum_{\bb\sigma \in \Sigma_N} \exp\Big(\frac{\beta N}2 m_{\bb\sigma}^2 +\frac{\beta N}{2}(Y_\mu^{(s)}(\bb\sigma,x))^2+\frac{{\beta N}}{2}\sum_{\nu \neq \mu}(Y_\nu^{(s)}(\bb\sigma))^2 \Big) }=\\
    &=\mathbb E_{\eta^\mu_i}\frac{\sum_{\bb\sigma \in \Sigma_N}( Y_\mu^{(s)}(\bb\sigma,x))  ^4 \exp\Big(\frac{\beta N}2 m_{\bb\sigma}^2 +{\beta \sqrt s}(x-\eta^\mu_i )\sigma_i Y^{(s)}_\mu(\bb\sigma)
    +\frac{{\beta N}}{2}\sum_{\nu}(Y_\nu^{(s)}(\bb\sigma))^2 \Big) 
    }{\sum_{\bb\sigma \in \Sigma_N} \exp\Big(\frac{\beta N}2 m_{\bb\sigma}^2 +{\beta \sqrt s}(x-\eta^\mu_i )\sigma_i Y^{(s)}_\mu(\bb\sigma)+\frac{{\beta N}}{2}\sum_{\nu}(Y_\nu^{(s)}(\bb\sigma))^2 \Big) }.
\end{ess}
Here, we included the average w.r.t. $\mathbb E_{\eta^\mu_i}$ (notice that the argument is independent of $\eta^\mu_i$) since we want to express the expectation as a perturbation of the usual Boltzmann average. Also, we dropped the terms independent of $\bb\sigma$, as they cancel out in the fraction. Finally, we included $(Y_\mu^{(s)}(\bb\sigma))^2$ in the sum. This implies that:
$$
\omega^{(x)}_{N,s}(( Y_\mu^{(s)}(\bb\sigma,x))  ^4)=\mathbb E_{\eta^\mu_i}\frac{\omega_{N,s}\Big(( Y_\mu^{(s)}(\bb\sigma,x))  ^4\exp\big({\beta \sqrt s}(x-\eta^\mu_i )\sigma_i Y^{(s)}_\mu(\bb\sigma)\big)\Big)}{\omega_{N,s}\Big(\exp\big({\beta \sqrt s}(x-\eta^\mu_i )\sigma_i Y^{(s)}_\mu(\bb\sigma)\big)\Big)}.
$$
Now
$$
\vert (x-\eta^\mu_i )\sigma_i Y^{(s)}_\mu(\bb\sigma) \vert \le \vert x-\eta^\mu_i \vert\cdot \vert Y_\mu^{(s)}(\bb\sigma)\vert \le 2 \vert Y_\mu^{(s)}(\bb\sigma)\vert,
$$
since the first factor is bounded as $\vert x-\eta^\mu_i\vert\le 2$, as $x\in [-1,1]$ and $\eta^\mu_i = \pm 1$. On the other hand:
\begin{ess}
    \vert Y_\mu^{(s)}(\bb\sigma)\vert =\vert \frac{\sqrt s}{N}\sum_i \eta^\mu_i \sigma_i + \frac{\sqrt{1-s}}{N}\sum_{i}\xi^\mu_i \sigma_i \vert \le \sqrt s + \sqrt{1-s}\frac 1 N\sum_{i=1}^N \vert \xi^\mu_i \vert,
\end{ess}
which is $\bb\sigma$-independent. We call $S_\mu= \frac 1 N\sum_{i=1}^N \vert \xi^\mu_i \vert \in [0,\infty)$, whose expectation is $\mathbb E_{\bb\xi} S_{\mu} = \sqrt{2/\pi}$. Then, we have
$$
\vert (x-\eta^\mu_i )\sigma_i Y^{(s)}_\mu(\bb\sigma) \vert\le 2(\sqrt s +\sqrt{1-s}S_\mu),
$$
which in turn implies
$$
\omega^{(x)}_{N,s}(( Y_\mu^{(s)}(\bb\sigma,x))  ^4)\le \exp(4\beta \sqrt s (\sqrt s+\sqrt {1-s} S_\mu))\mathbb E_{\eta^\mu_i}\omega _{N,s}(( Y_\mu^{(s)}(\bb\sigma,x))  ^4).
$$
Using again Eq. \eqref{eq:why}, we have
\begin{ess}
\omega _{N,s}(( Y_\mu^{(s)}(\bb\sigma,x))  ^4 )&=\omega _{N,s}( ( Y_\mu^{(s)}(\bb\sigma))  ^4)+ \sum_{k=1}^4 \frac{s^{k/2}}{N^k}\binom{4}{k} \omega_{N,s}\Big(( Y_\mu^{(s)}(\bb\sigma))^{4-k} \big((x-\eta^\mu_i) \sigma_i\big)^k\Big)    \le\\
&\le  \omega _{N,s}( ( Y_\mu^{(s)}(\bb\sigma))  ^4)+ \sum_{k=1}^4 2^k\frac{s^{k/2}}{N^k}\binom{4}{k} \omega_{N,s}(\vert Y_\mu^{(s)}(\bb\sigma)\vert ^{4-k}).
\end{ess}
Again by H\"older inequality, $\omega_{N,s}(\vert Y_\mu^{(s)}(\bb\sigma)\vert ^{4-k}) \le \omega_{N,s}(( Y_\mu^{(s)}(\bb\sigma)) ^{4})^{\frac{4-k}4}$ for $k=1,2,3$. Then
\begin{es}
  \omega _{N,s}(( Y_\mu^{(s)}(\bb\sigma,x))  ^4 )&\le  \omega _{N,s}( ( Y_\mu^{(s)}(\bb\sigma))  ^4)+ \sum_{k=1}^4 \frac{(4s)^{k/2}}{N^k}\binom{4}{k}\omega_{N,s}( Y_\mu^{(s)}(\bb\sigma) ^{4})^{\frac{4-k}4} .
\end{es}
This is now independent of $x$, thus
\begin{es}\label{eq:annoying_formula}
    \mathbb E_{\bb\eta_{\neg \{i,\mu\},\bb\xi}}\sup_{\vert x\vert \le 1}\omega_{N,s}^{(x)}(( Y_\mu^{(s)}(\bb\sigma,x))  ^4)&\le\mathbb E_{\bb\xi}\Big\{\exp(4\beta \sqrt s (\sqrt s+\sqrt {1-s} S_\mu))\Big[\mathbb E_{\bb\eta}\omega _{N,s}( ( Y_\mu^{(s)}(\bb\sigma))  ^4)\\&+ \sum_{k=1}^4 \frac{(4s)^{k/2}}{N^k}\binom{4}{k}\mathbb E_{\bb\eta}\omega_{N,s}( Y_\mu^{(s)}(\bb\sigma) ^{4})^{\frac{4-k}4}\Big]\Big\}.
\end{es}
Since $x\to x^\theta$, $\theta = {\tfrac{4-k}{4}}$, is concave for $k=1,2,3$, by Jensen's inequality:
\begin{ess}
\mathbb E_{\bb\eta_{\neg \{i,\mu\},\bb\xi}}\sup_{\vert x\vert \le 1}\omega_{N,s}^{(x)}(( Y_\mu^{(s)}(\bb\sigma,x))  ^4)&\le e^{4\beta s+4\beta \sqrt {\tfrac{2 s(1-s)}{\pi}} }\mathbb E_{\bb\xi}\Big\{\exp(4\beta \sqrt {s(1-s)} \tilde S_\mu)\cdot \\ &\cdot \Big[\mathbb E_{\bb\eta}\omega _{N,s}( ( Y_\mu^{(s)}(\bb\sigma))  ^4)+ \sum_{k=1}^4 \frac{(4s)^{k/2}}{N^k}\binom{4}{k}[\mathbb E_{\bb\eta}\omega_{N,s}( Y_\mu^{(s)}(\bb\sigma) ^{4})]^{\frac{4-k}4}\Big]    \Big\}.
\end{ess}
where we defined $\tilde S_\mu = S_\mu -\mathbb E_{\bb\xi} S_\mu$ and used the fact that $\mathbb E_{\bb\xi} S_\mu= \sqrt{\tfrac 2\pi}$. The presence of $\tilde S_\mu$ prevents us from directly transferring the expectation $\mathbb E_{\bb \xi}$ on the relevant Boltzmann averages. We thus proceed by truncating $\tilde S_\mu $ below and above a cut $u_0>0$, i.e. splitting the expectation in r.h.s. of Eq. \eqref{eq:annoying_formula} as $\mathbb E_{\bb\xi} 1_{\tilde S_\mu <u_0}(\dots)+\mathbb E_{\bb\xi} 1_{\tilde S_\mu \ge u_0}(\dots)$. In the first case, we have
\begin{ess}
    &\mathbb E_{\bb\xi}\Big\{1_{\tilde S_\mu <u_0}\exp(4\beta \sqrt {s(1-s)} \tilde S_\mu) \Big[\mathbb E_{\bb\eta}\omega _{N,s}( ( Y_\mu^{(s)}(\bb\sigma))  ^4)+ \sum_{k=1}^4 \frac{(4s)^{k/2}}{N^k}\binom{4}{k}[\mathbb E_{\bb\eta}\omega_{N,s}( Y_\mu^{(s)}(\bb\sigma) ^{4})]^{\frac{4-k}4}\Big]    \Big\}\le\\
 \le \,  &\exp(4\beta \sqrt {s(1-s)}u_0)\mathbb E_{\bb\xi}\Big\{1_{\tilde S_\mu <u_0}\Big[\mathbb E_{\bb\eta}\omega _{N,s}( ( Y_\mu^{(s)}(\bb\sigma))  ^4)+ \sum_{k=1}^4 \frac{(4s)^{k/2}}{N^k}\binom{4}{k}[\mathbb E_{\bb\eta}\omega_{N,s}( Y_\mu^{(s)}(\bb\sigma) ^{4})]^{\frac{4-k}4}\Big]    \Big\}\le
 \\
 \le  \, &\exp(4\beta \sqrt {s(1-s)}u_0)\Big[\mathbb E_{\bb\eta,\bb\xi}\omega _{N,s}( ( Y_\mu^{(s)}(\bb\sigma))  ^4)+ \sum_{k=1}^4 \frac{(4s)^{k/2}}{N^k}\binom{4}{k}[\mathbb E_{\bb\eta,\bb\xi}\omega_{N,s}( Y_\mu^{(s)}(\bb\sigma) ^{4})]^{\frac{4-k}4}\Big],
\end{ess}
where in the last line we used $1_{\tilde S_\mu <u_0}\le 1$ and again Jensen's inequality on the concave map $x\to x^\theta$, $\theta = {\tfrac{4-k}{4}}$ for $k=1,2,3$. If the fields $Y_\mu^{(t)}$ are non-condensed, $\mathbb E_{\bb\eta,\bb\xi}\omega_{N,s}( Y_\mu^{(s)}(\bb\sigma) ^{4})\le \frac {C(\alpha,\beta)}{N^2}$, we can bound the first contribution as
\begin{es}
  \mathbb E_{\bb\xi} 1_{\tilde S_\mu <u_0}(\dots)&\le \exp(4\beta \sqrt {s(1-s)}u_0)  \Big[\frac{C(\alpha,\beta)}{N^2}+ \sum_{k=1}^4 \frac{(4s)^{k/2}}{N^{2+\frac k2}}\binom{4}{k}C(\alpha,\beta)^{\frac{4-k}4}\Big] =\\
  &= \frac1{N^2}\exp(4\beta \sqrt {s(1-s)}u_0) \Big(\sqrt{\frac{4s}{N}}+C(\alpha,\beta )^{\frac14}\Big)^4.
\end{es}
For the complementary event, we need tail conditions on the random variable $\tilde S_\mu$.
First, by the Chernoff bound, for all $u>0$ one has
$$
\mathbb P(\tilde S_\mu \ge u)\le \inf_{\lambda>0} \Big[ \exp(-N u\lambda) M_{N\tilde S_\mu}(\lambda)\Big],
$$
with $M$ being the moment generating function of $N\tilde S_\mu$. By independence of $\xi^\mu_i$:
$$
M_{N\tilde S_\mu}(\lambda)= \mathbb E_{\bb\xi^\mu}\exp\Big(\lambda \sum_i( \vert \xi_i^\mu\vert  - \sqrt{\tfrac 2\pi})\Big)=\Big( \mathbb E_{\xi^\mu_i}\exp\big(\lambda (\vert \xi^\mu_i\vert  - \sqrt{\tfrac 2\pi})\big)\Big)^N .
$$
Since $\xi_i^\mu \sim \mc N(0,1)$, we have $\mathbb E_{\xi^\mu}\exp\big(\lambda \vert \xi^\mu\vert \big)= 2 \exp\big(\tfrac{\lambda^2}2\big)\Phi(\lambda)$. Using the fact that $\Phi(\lambda)\le \frac12 e^{\lambda \sqrt{\frac 2 \pi}}$ (following from the concavity of $g(\lambda)=\log (2\Phi(\lambda))$, which guarantees that $g(\lambda)\le g(0)+\lambda g'(0)= \lambda \sqrt{2/\pi}$), we have $\mathbb E_{\xi^\mu}\exp\big(\lambda \vert \xi^\mu\vert \big)\le \exp(\frac{\lambda^2}{2}+\lambda \sqrt{\frac 2 \pi})$. Using this result, we have
$$
\mathbb P(\tilde S_\mu \ge u)\le \inf_{\lambda>0}  \exp\big(-N u \lambda +\tfrac{\lambda ^2 N}{2}\big)   = \exp\big(-\tfrac{N u^2}{2}\big).
$$
The second ingredient is again $\vert Y_\mu^{(s)}\vert \le \sqrt s +\sqrt{1-s} S_\mu $ (which is independent of $\bb\sigma$), which in turn implies
$$
\omega_{N,s}((Y_\mu^{(s)}(\bb\sigma))^4)\le  [\sqrt s +\sqrt{\tfrac{2(1-s)}\pi} +\sqrt{1-s}\tilde S_\mu]^4.
$$
From this, it follows that
\begin{ess}
    &\mathbb E_{\bb\xi}\Big\{1_{\tilde S_\mu \ge u_0}\exp(4\beta \sqrt {s(1-s)} \tilde S_\mu) \Big[\mathbb E_{\bb\eta}\omega _{N,s}( ( Y_\mu^{(s)}(\bb\sigma))  ^4)+ \sum_{k=1}^4 \frac{(4s)^{k/2}}{N^k}\binom{4}{k}[\mathbb E_{\bb\eta}\omega_{N,s}( Y_\mu^{(s)}(\bb\sigma) ^{4})]^{\frac{4-k}4}\Big]    \Big\}\le
   \\ \le \, & \mathbb E_{\bb\xi}\Big\{1_{\tilde S_\mu \ge u_0}\exp(4\beta \sqrt {s(1-s)} \tilde S_\mu) \Big[ [\sqrt s +\sqrt{\tfrac{2(1-s)}\pi} +\sqrt{1-s}\tilde S_\mu]^4+ \sum_{k=1}^4 \binom{4}{k}\Big(\sqrt{\frac{4s}{N^2}}\Big)^k[ \sqrt s +\sqrt{\tfrac{2(1-s)}\pi} +\sqrt{1-s}\tilde S_\mu]^{{4-k}}\Big]    \Big\}=\\
   =\,&\mathbb E_{\bb\xi}\Big\{1_{\tilde S_\mu \ge u_0}\exp(4\beta \sqrt {s(1-s)} \tilde S_\mu) \Big(\sqrt{\frac{4s}{N^2}}+ \sqrt s +\sqrt{\tfrac{2(1-s)}\pi} +\sqrt{1-s}\tilde S_\mu\Big)^4\Big\}.
\end{ess}
The second contribution thus has the form of $\mathbb E [1_{\tilde S_\mu \ge u_0} \phi(\tilde S_\mu)]$ with $\phi(x) = (a+b x)^4 \exp(c x)$ and
\begin{eqnarray*}
    a&=& \sqrt s+ \sqrt{\frac{2(1-s)}{\pi}},\\
    b&=& \sqrt {1-s},\\
    c &=&4 \beta \sqrt{s(1-s)}.
\end{eqnarray*}
We use the formula
$$
\mathbb E [1_{\tilde S_\mu \ge u_0} \phi(\tilde S_\mu)] = \phi(u_0) \mathbb P(\tilde S_\mu \ge u_0)+\int_{u_0}^\infty du \phi'(u) \mathbb P (\tilde S_\mu \ge u).
$$
From the tail behavior of the distribution of $\tilde S_\mu$, we have
$$
\mathbb E [1_{\tilde S_\mu \ge u_0} \phi(\tilde S_\mu)]\le \phi(u_0) \exp(-\tfrac{N u_0 ^2}{2})+\int _{u_0}^\infty du \phi'(u) \exp(-\tfrac{N u ^2}{2}).
$$
Notice that $\phi'(u) = [c (a+b u)^4+4b(a+bu)^3]\exp(c u)$, so that the second contribution can be written in terms of integrals of the form $I_k(u_0)=\int _{u_0}^\infty du u^k \exp(cu -\tfrac{N u^2}{2})$. In particular, we can use the inequality $(a+b u)^n\le 2^{n-1} (a^n+b^n u^n)$ for any positive integer $n$, and bound everything in terms of these integrals:
$$
\int _{u_0}^\infty du \phi'(u) \exp(-\tfrac{N u ^2}{2})\le 8(ca^4+2b a^3)I_0(u_0)+16 b^4 I_3(u_0) + 8c b^4 I_4(u_0).
$$
Now with a simple change of variable we have
$$
I_k (u_0)= e^{\tfrac{c^2}{2N}}\sum_{l=0}^k \binom{k}{l}\Big(\frac{c}{N}\Big)^{k-l}\int _\delta ^\infty du u^l \exp(-\tfrac{N u^2}{2}),
$$
with $\delta = u_0-\tfrac cN$. Let us fix $N_0 = 2[\sup_{s\in [0,1]}c]/u_0 = 4\beta/u_0$; then, for $N\ge N_0$, $\delta\ge \frac{u_0}2>0$. We call $J_l (\delta)$ the integrals on the r.h.s. Integrating by parts, it is easy to show that they satisfy the recursion relation
$$
J_l (\delta) = \frac{\delta^{l-1}}{N}\exp(-\tfrac{N\delta^2}{2})+\frac{l-1}{N}J_{l-2}(\delta),
$$
together with the simple bound $J_0 (\delta) \le \frac1{N\delta}\exp(-\tfrac{N\delta^2}{2})$ (following from Mills' inequality) and $J_1(\delta) = \frac1N \exp(-\tfrac{N\delta^2}{2})$. Iterating the recursion, one gets in particular
\begin{ess}
    J_3(\delta) = e^{-\frac{N\delta^2}{2}}\Big(\frac{\delta^2}{N}+\frac{2}{N^2}\Big),
    \qquad
    J_4(\delta) \le e^{-\frac{N\delta^2}{2}}\Big(\frac{\delta^3}{N}+\frac{3\delta}{N^2}+\frac{3}{N^3\delta}\Big),
\end{ess}
so that, in general, $J_l(\delta) \le \frac1N \kappa_l^{(N)}(\delta) \exp(-\tfrac{N \delta^2}{2})$. Since $\frac{u_0}2\le \delta \le u_0$ for $N\ge N_0$, the coefficients $\kappa_l^{(N)}(\delta)$ are bounded uniformly in $N\ge N_0$ by explicit constants depending on $u_0$ only. With these bounds, we can thus write
$$
I_k (u_0) \le \frac {e^{\tfrac{c^2}{2N}-\tfrac{N \delta^2}{2}}}N \sum_{l=0}^k \binom{k}{l}\Big(\frac{c}{N}\Big)^{k-l}\kappa_l^{(N)} (\delta)=\frac {e^{\tfrac{c^2}{2N}-\tfrac{N \delta^2}{2}}}N \zeta_k^{(N)} (c,u_0),
$$
with $\zeta_k^{(N)} (c,u_0)=\sum_{l=0}^k \binom{k}{l}(\frac{c}{N})^{k-l}\kappa_l ^{(N)}(\delta)$, bounded uniformly in $N\ge N_0$. Then, for $N\ge N_0$:
$$
\int_{u_0}^\infty du \phi'(u)\le  \frac {e^{\tfrac{c^2}{2N}-\tfrac{N \delta^2}{2}}}N\big(8(ca^4+2b a^3)\zeta_0^{(N)}(c,u_0)+16 b^4 \zeta_3^{(N)}(c,u_0) + 8c b^4 \zeta_4^{(N)}(c,u_0)\big)\le \frac {e^{\tfrac{c^2}{2N}-\tfrac{N \delta^2}{2}}}N C_{\beta,s},
$$
for some finite constant $C_{\beta,s}$ independent of $N$ (the terms inside the round brackets being uniformly bounded for $N\ge N_0$, it suffices to take the supremum over $N\ge N_0$). Then, using $\delta\ge \frac{u_0}2$, the tail contribution is globally
\begin{es}
      &\mathbb E_{\bb\xi}\Big\{1_{\tilde S_\mu \ge u_0}\exp(4\beta \sqrt {s(1-s)} \tilde S_\mu) \Big[\mathbb E_{\bb\eta}\omega _{N,s}( ( Y_\mu^{(s)}(\bb\sigma))  ^4)\\&+ \sum_{k=1}^4 \frac{(4s)^{k/2}}{N^k}\binom{4}{k}[\mathbb E_{\bb\eta}\omega_{N,s}( Y_\mu^{(s)}(\bb\sigma) ^{4})]^{\frac{4-k}4}\Big]    \Big\}\le \phi(u_0) e^{-\tfrac{N u_0 ^2}{2}}+\frac {e^{\tfrac{c^2}{2N}-\tfrac{N u_0^2}{8}}}N C_{\beta,s}.
\end{es}
Merging the two events $\tilde S_\mu <u_0 $  and $\tilde S_\mu \ge u_0$, we finally obtain the bound
\begin{ess}
  \mathbb E_{\bb\eta_{\neg \{i,\mu\},\bb\xi}}\sup_{\vert x\vert \le 1}\omega_{N,s}^{(x)}(( Y_\mu^{(s)}(\bb\sigma,x))  ^4)\le \frac{e^{4\beta s+4\beta \sqrt {\tfrac{2 s(1-s)}{\pi}} }}{N^2}&\Big[ \exp(4\beta \sqrt {s(1-s)}u_0) \Big(\sqrt{\frac{4s}{N}}+C(\alpha,\beta )^{\frac14}\Big)^4\\
  &+ \phi(u_0) N^2 e^{-\tfrac{N u_0 ^2}{2}}+N {e^{\tfrac{8\beta^2 s(1-s)}{N}-\tfrac{N u_0^2}{8}}} C_{\beta,s}
  \Big] ,
\end{ess}
holding for all $\mu=1,\dots,K$, for all $u_0>0$ and $N\ge N_0$.
Recalling that $$\vert\partial_s A_{N,K}(s)\vert\le \frac{\beta}{3N\sqrt s}\sum_{i,\mu}^{N,K}\mathbb E_{\bb \eta_{\neg \{i,\mu\}}, \bb \xi}\lVert f''' \lVert_{\infty,1},$$ and the inequality in Eq. \eqref{eq:3rd_der}, we have
\begin{ess}
    \vert\partial_s A_{N,K}(s)\vert\le \frac{26 \alpha_N\beta ^4 s e^{4\beta s+4\beta \sqrt {\tfrac{2 s(1-s)}{\pi}} }}{3N}&\Big[ \exp(4\beta \sqrt {s(1-s)}u_0) \Big(\sqrt{\frac{4s}{N}}+C(\alpha,\beta )^{\frac14}\Big)^4\\
  &+ \phi(u_0) N^2 e^{-\tfrac{N u_0 ^2}{2}}+N {e^{\tfrac{8\beta^2 s(1-s)}{ N}-\tfrac{N u_0^2}{8}}} C_{\beta,s}
  \Big] .
\end{ess}
Clearly, $\lim_{N\to\infty} \vert\partial_s A_{N,K}(s)\vert=0$ for all $s\in [0,1]$. Also, since $0\le s(1-s)\le \tfrac14$ for $s\in[0,1]$, $\alpha_N \le \alpha$ for $N>\alpha^{-1}$, and $\sup_{s\in [0,1]}C_{\beta,s}<+\infty$ since $a\le 1+\sqrt{2/\pi}$, $b \le 1$ and $c \le 2\beta$ for all $s\in [0,1]$, we can eventually bound the modulus of the derivative of the interpolating pressure by a constant $\mc C$ reading
\begin{ess}
\mc  C = \sup_{s\in [0,1]}\sup_{N\ge \max\{\alpha^{-1},N_0\}}\frac{26 \alpha \beta ^4 s e^{4\beta s+4\beta \sqrt {\tfrac{2 s(1-s)}{\pi}} }}{3N}&\Big[ \exp(4\beta \sqrt {s(1-s)}u_0) \Big(\sqrt{\frac{4s}{N}}+C(\alpha,\beta )^{\frac14}\Big)^4\\&+ \phi(u_0) N^2 e^{-\tfrac{N u_0 ^2}{2}}+N {e^{\tfrac{8\beta^2 s(1-s)}{ N}-\tfrac{N u_0^2}{8}}} C_{\beta,s}
  \Big].
\end{ess}
By the dominated convergence theorem
$$
\lim_{N\to\infty}\int_0 ^1 ds\,  \vert\partial_s A_{N,K}(s)\vert= \int_0 ^1 ds\, \lim_{N\to\infty} \vert\partial_s A_{N,K}(s)\vert=0.
$$
This finally implies, by inequality \eqref{bound_universality}
$$
\lim_{N\to\infty}\vert A^{(H)}_{N,K}(\beta)- A_{N,K}(\beta)\vert=0,
$$
thus proving our claim.
\end{proof}
We conclude by stressing that, by virtue of Prop. \ref{prp:universality_ergo}, universality holds unconditionally for $\beta <1$ (regardless of the value of the load $\alpha$). For $\beta\ge 1$, the result is conditioned on the (physically motivated) assumption of non-condensation of the bulk of noisy patterns.
}

\bibliographystyle{unsrt}   

\end{document}